\documentclass[aps,pra,preprint,superscriptaddress]{revtex4-2}

\usepackage{amsmath,amssymb}
\usepackage{mathrsfs}
\usepackage{graphicx}
\usepackage{epstopdf}
\usepackage{bm,color,bbm}
\usepackage{natbib}
\usepackage{hyperref}
\usepackage{caption}
\usepackage{fancyhdr}
\usepackage{multirow} 

\newcommand{\ket}[1]{ |{#1} \rangle}

\newcommand{\an}[1]{\left\langle{#1}\right\rangle}

\begin{document}
\fancyhead[R]{\ifnum\value{page}<2\relax\else\thepage\fi}

\title{Frequency auto-homogenization using group-velocity-matched downconversion}

\author{Dylan Heberle}
\email[]{dylan.heberle@techngs.com}
\affiliation{School of Applied and Engineering Physics, Cornell University, Ithaca, New York, 14853, USA}
\affiliation{Griffiss Institute, Rome, New York, 13441, USA} 
\affiliation{Technergetics, LLC, Utica, New York, 13502, USA} 
\author{Christopher C. Tison}
\author{James Schneeloch}
\author{A. Matthew Smith}
\author{Paul M. Alsing}
\affiliation{Air Force Research Laboratory, Information Directorate, Rome, New York, 13441, USA}
\author{Jeffrey Moses}
\affiliation{School of Applied and Engineering Physics, Cornell University, Ithaca, New York, 14853, USA}
\author{Michael L. Fanto}
\affiliation{Air Force Research Laboratory, Information Directorate, Rome, New York, 13441, USA}

\date{\today}

\begin{abstract}

With the stability of integrated photonics at network nodes and the advantages of photons as flying qubits, photonic quantum information processing (PQIP) makes quantum networks increasingly scalable. However, scaling up PQIP requires the preparation of many identical single photons which is limited by the spectral distinguishability of integrated single-photon sources due to variations in fabrication or local environment. To address this, we introduce frequency auto-homogenization via group-velocity-matched downconversion to remove spectral distinguishability in varying quantum emitters. We present our theory using $\chi^{(2)}$ quantum frequency conversion and show proof-of-principle data in a free-space optical setup.

\end{abstract}

\maketitle

\section{Introduction}

Quantum information science relies on the availability of indistinguishable single-photon sources for state preparation and quantum computation. Optical implementations of quantum logic gates such as the controlled-NOT gate \cite{ralph_2001}, which is the quantum equivalent of the classical XOR gate, and the type-II fusion gate \cite{browne_2005}, which projects the incident two-photon state onto an even- or odd-parity subspace, depend on the indistinguishability of the incident photons. For example, the error in the controlled-NOT gate exceeds the fault-tolerant threshold of $1\%$ for linear optical quantum computing \cite{knill_2004} when spectral distinguishability is more than $10\%$ of the photon bandwidth ($1/e^2$ half width) \cite{rohde_2005}, and the error in the type-II fusion gate exceeds the fault-tolerant threshold of $0.01\%$ for optical cluster-state quantum computing \cite{dawson_2006} when the distinguishability is more than $\sim$2$\%$ of the photon bandwidth ($1/e^2$ half width) \cite{rohde_2006}. Thus, the center wavelengths of two single-photon emitters with 1-nm bandwidths must vary by $<$0.1 nm for the controlled-NOT gate and by $<$0.02 nm for  the type-II fusion gate. Moreover, the interference of partially distinguishable bosons has been studied in the context of multimode networks and boson sampling \cite{shchesnovich_2014, shchesnovich_2015, tichy_2014, tichy_2015, rohde_2015, shi_2022}. As distinguishability increases, the probability of measuring a given output state cannot be described as a simple interpolation between the probability resulting from completely distinguishable particles and that from indistinguishable particles \cite{tichy_2014, tichy_2015}. In the case of boson sampling, distinguishability reduces computational complexity requiring more photons and modes to demonstrate increased performance over classical alternatives \cite{shchesnovich_2014, shchesnovich_2015, tichy_2015, rohde_2015, shi_2022}. Recently, Zhong et al. \cite{zhong_2020} performed Gaussian boson sampling with 50 single-mode squeezed states in a 100-mode interferometer in which great care had to be taken to eliminate any source of distinghishability including frequency---whereby the wavelength detuning of the 12-nm FWHM photons was kept $<$1 nm using PID temperature control. Finally, long-distance quantum communication in fiber requires indistinguishable single-photon sources in the infrared due to the wavelength requirements of the global telecommunications infrastructure \cite{wehner_2018}.

Solid-state devices such as quantum dots and color centers constitute promising avenues for bright, deterministic single-photon sources \cite{aharonovich_2016}. 
However, these devices often emit in the visible region of the electromagnetic spectrum and suffer from inhomogeneous spectral broadening and time-dependent emission, severely reducing the yield of identical emitters \cite{Ding_2016,Sipahigil_2012}. Fortunately, the center frequency of these sources can be easily translated to telecommunication bands while preserving their quantum state using quantum frequency conversion such as three- and four-wave mixing in $\chi^{(2)}$ and $\chi^{(3)}$ nonlinear materials respectively \cite{kumar_1990, ikuta_2011, zaske_2012, degreve_2012, albrecht_2014, li_2016, dreau2018quantum, bock2018high, da2022pure}. Spectral distinguishability of single-photon sources has been addressed through direct manipulation via temperature tuning \cite{kiraz_2001, faraon_2007}, stress tuning \cite{seidl_2006}, Stark shift \cite{findeis_2001}, and laser annealing \cite{dubowski_2018} and through frequency translation via nonlinear wave interactions such as active feed-forward spectral multiplexing \cite{yu_2021, puigibert_2017}, cross phase modulation \cite{matsuda_2016}, and three- \cite{takesue_2008, ates_2012, weber_2019} or four-wave mixing \cite{singh_2019}. The direct methods used to manipulate the spectral emission often result in optical shifts on the order of a few hundred GHz or less and require isolating each source for independent control. Alternatively, the frequency translation methods allow for larger shifts; however, a priori knowledge of the photon emission spectrum or feed-forward control is required, limiting device scalability and increasing complexity. 

In addition to frequency translation, quantum frequency conversion via three-wave mixing has been used to construct a quantum pulse gate (QPG) and a quantum pulse shaper (QPS) which allow manipulation of the temporal-mode structure of ultrafast quantum sources \cite{eckstein_2011, brecht_2011}. The QPG selects a single temporal mode from a multimode input using sum frequency generation with a shaped pump pulse, and the QPS generates a target output pulse mode from a Gaussian-shaped input spectrum using difference frequency generation (DFG) with a shaped pump pulse \cite{eckstein_2011, brecht_2011, brecht_2014}. Because these temporal modes span an infinite-dimensional Hilbert space and are compatible with waveguide devices and fiber systems, they have been proposed as a framework for quantum information science \cite{brecht_2015}.

Here, we propose a device that converts mismatched spectral modes from diverse quantum emitters to a single output spectral mode at 1.55 \textmu m, erasing spectral distinguishability. The device uses group-velocity-matched downconversion with a broadband pump. Because no a priori information about the input photons from the inhomogenous quantum emitters is required, we call this process frequency \emph{auto}-homogenization. Instead of targeting one specific mode as in the QPG and QPS, the broadband pump field defines a spectral window within which any input spectrum will be converted to a single output spectral mode, with the efficiency determined by the overlap between the fundamental Schmidt mode of the homogenization device and the input spectrum. We emphasize our device not only translates all the emitted photons' center frequencies to 1.55 \textmu m but also auto-homogenizes the output spectral mode.

In the following sections, we review quantum frequency downconversion to calculate the output spectral distribution of our device; we show that a joint coupling amplitude (JCA) determines how the input states are mapped to the output states; and, through the JCA, we determine the requirements for frequency auto-homogenization. Finally, we present results from analytic calculations, split-step Fourier beam propagation simulations, and measurements of the joint coupling intensity (JCI) of Rb:KTP waveguide devices as proof-of-principle measurements of frequency auto-homogenization.

\section{Theory}

Our frequency auto-homogenization device converts an input photon of angular frequency $\omega_{i}$ derived from a broad, heterogeneous spectral distribution to a spectrally homogeneous output photon of angular frequency $\omega_{o}$ in the presence of a classically bright pump field of angular frequency $\omega_{p}$. This process is described by the unitary time-evolution operator 
\begin{equation} \label{eq:TimeIntegratedHamiltonian}
    \hat{U} \approx \exp{ \left( \mathcal{C} \int d\omega_i d\omega_o f(\omega_i,\omega_o) \hat{a}(\omega_i) \hat{b}^\dagger(\omega_o) + H.C. \right)}
\end{equation}
for collinear DFG where $\hat{a}(\omega_i)$ and $\hat{b}^\dagger(\omega_o)$ are the respective annihilation and creation operators for photons in the input and output modes at their respective frequencies, $f(\omega_i,\omega_o)$ is the normalized JCA between the input and output spectral modes, $\mathcal{C}$ is an evolution parameter describing the strength of the interaction, and $H.C.$ is the Hermitian conjugate. The complete derivation is included in Appendix \ref{app:HintJCA}, and an analogous time-evolution operator for pulse-pumped spontaneous parametric downconversion is derived in \cite{grice_1997}.

To determine how the output fields vary for each input field, we consider input and output fields of single-photon Fock states given by 
\begin{align}
    \ket{\psi_i} &= \int d\omega_i \alpha^*(\omega_i) \hat{a}^\dagger(\omega_i) \ket{0, 0} \label{eq:inputstate} \\
    \ket{\psi_o} &= \int d\omega_o \beta^*(\omega_o) \hat{b}^\dagger(\omega_o) \ket{0, 0} \label{eq:outputstate}
\end{align}
respectively. The conjugated input and output spectral amplitudes are $\alpha^*(\omega_i)$ and $\beta^*(\omega_o)$. Note, each input field could have a different carrier frequency or spectral distribution depending on the inhomogeneity of the photon emitter. After frequency conversion the total field is given by
\begin{align}
    \ket{\psi} &= \hat{U}(0, t)\ket{\psi_i} \nonumber \\
    \ket{\psi} &\approx \exp{\left[ \mathcal{C} \int d\omega_i d\omega_o f(\omega_i, \omega_o) \hat{a}(\omega_i) \hat{b}^\dagger(\omega_o) + H.C. \right]} \ket{\psi_i} \label{eq:totalfield1}
\end{align}
where we have ignored the effects of time ordering as the conversion process of interest must be stimulated by the presence of an input field \cite{quesada2014effects}. Substituting Eq. \ref{eq:inputstate} into Eq. \ref{eq:totalfield1} yields
\begin{equation}
    \ket{\psi} = \int d\omega_i \alpha^*(\omega_i) \hat{a}^\dagger(\omega_i) \ket{0, 0} + \mathcal{C}\int d\omega_i d\omega_o f(\omega_i, \omega_o) \alpha^*(\omega_i) \hat{b}^\dagger(\omega_o) \ket{0, 0} + \mathcal{O}(\mathcal{C}^2). \label{eq:totalfield2}
\end{equation}
Upon inspection, Eq. \ref{eq:totalfield2} can be expressed as
\begin{equation}
    \ket{\psi} = \ket{\psi_i} + \mathcal{C} \ket{\psi_o} + \mathcal{O}(\mathcal{C}^2)
\end{equation}
where we find
\begin{equation}
    \beta^*(\omega_o) = \int d\omega_i f(\omega_i, \omega_o) \alpha^*(\omega_i).
\end{equation}

Notably, the JCA describes how the input modes are mapped to the output modes. Following Appendix \ref{app:HintJCA}, the normalized JCA is given by
\begin{equation} \label{eq:JCA}
    f(\omega_i, \omega_o) = s^*(\omega_i - \omega_o) \phi(\omega_i, \omega_o)
\end{equation}
where $s^*(\omega_i - \omega_o)$ is the conjugated pump spectral amplitude and $\phi(\omega_i, \omega_o)$ is the phase-matching function (PMF). For collinear DFG, the PMF is given by
\begin{equation} \label{eq:PMF}
    \phi(\omega_i, \omega_o) = \frac{1}{\mathcal{N}}\operatorname{sinc}{\left( \frac{\Delta k L}{2} \right)}
\end{equation}
where $\mathcal{N}$ is a normalization coefficient, L is the length of the homogenization crystal, and $\Delta k = k_i - k_o - k_p$ is the wave-vector mismatch in which $k_i$, $k_o$, and $k_p$ are the wavenumbers of the input, output, and pump fields respectively. In the following subsections, we describe how the JCA can be engineered to achieve frequency auto-homogenization, explore the performance of the proposed device using modal analysis, and propose interferometric visibility as a metric to quantify device performance.

\subsection{Engineering the JCA}

\begin{figure*}[t]
\includegraphics[width=\textwidth]{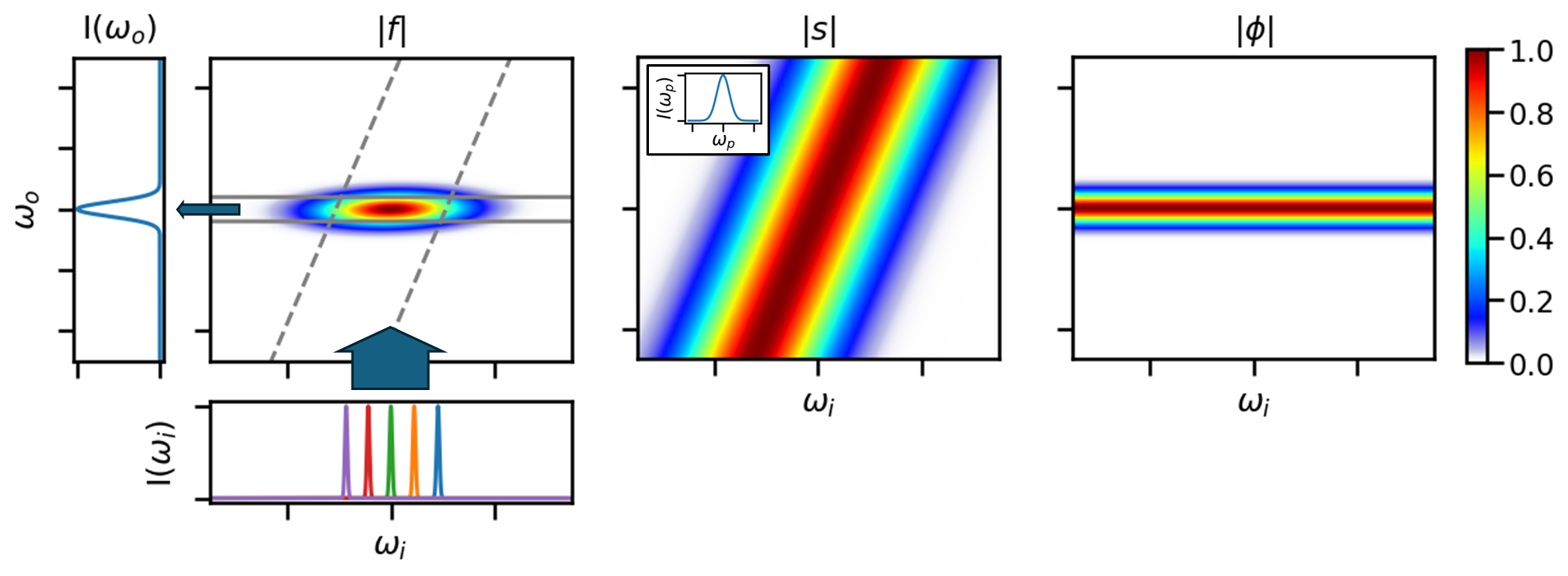}
\caption{Depiction of frequency auto-homogenization. (Left) Magnitude of a JCA ($f$) that maps distinguishable input spectra ($I(\omega_i)$, bottom) from an inhomogeneous spectral distribution to the same output spectral mode ($I(\omega_o)$). The dashed and solid gray lines indicate the FWHM values of the pump spectral amplitude ($s$) and PMF ($\phi$) respectively. (Center) Magnitude of the pump spectral amplitude as a function of input and output angular frequencies ($\omega_i$, $\omega_o$). The inset shows the pump spectrum as a function of $\omega_p$. (Right) Magnitude of a suitable PMF that enables efficient conversion of a broad input distribution to a single output frequency. \label{fig:JCAschematic}}
\end{figure*}

For frequency auto-homogenization, our goal is to convert a set of distinguishable input spectra derived from an inhomogeneously broadened spectral distribution to a single output spectral mode without a priori knowledge and individualized tuning as shown on the left side of Fig. \ref{fig:JCAschematic}. 
To accomplish this, our target JCA is an elongated horizontal function in the parameter space given by $\omega_i$ and $\omega_o$ that is wide along the axis corresponding to $\omega_i$ and narrow along the axis corresponding to $\omega_o$. From Eq. \ref{eq:JCA}, the JCA is determined by the pump spectral amplitude and the PMF which can be engineered to obtain the desired shape. First, the pump spectral amplitude must be broad enough to satisfy energy conservation between the target output wavelength and the entire input spectral distribution. This ensures that the JCA will be wide enough along $\omega_i$ to accommodate all input spectra. Because the pump spectral amplitude is constrained along the line dictated by energy conservation between the carrier angular frequencies, $\overline{\omega}_p=\overline{\omega}_\mathrm{i}-\overline{\omega}_o$, as shown by the center plot in Fig. \ref{fig:JCAschematic}, the PMF must restrict the width of the JCA along $\omega_o$ ensuring efficient conversion of a broad input distribution to a single output frequency. This is achieved when the PMF lies horizontally in the parameter space given by $\omega_i$ and $\omega_o$ as shown on the right side of Fig. \ref{fig:JCAschematic}. The orientation of the PMF can be determined by expanding the wavenumbers to first order around the carrier frequencies and solving for the wave-vector mismatch.

The expansion of the wavenumbers gives
\begin{equation}
k_j(\omega_j) = k_j(\overline{\omega}_j) +  v_j^{-1}\Omega_j,
\end{equation}
where $v_j^{-1}=\frac{\partial k_j}{\partial \omega_j}\Big|_{\overline{\omega}_j}$ is the inverse of the group velocity evaluated at the carrier angular frequency $\overline{\omega}_j$ and $\Omega_j=\omega_j - \overline{\omega}_j$ is the relative angular frequency. Solving for the wave-vector mismatch gives 
\begin{align}
\Delta k &= k_i - k_p - k_o \nonumber \\
&= k_i(\overline{\omega}_i) + v_i^{-1} \Omega_i -  k_p(\overline{\omega}_p) - v_p^{-1} \Omega_p -  k_o(\overline{\omega}_o) - v_o^{-1} \Omega_o.
\label{eq:CAwavevectormismatch}
\end{align}
Using $\Omega_p = \Omega_i - \Omega_o$, Eq. \ref{eq:CAwavevectormismatch} can be simplified to
\begin{equation}
\Delta k = \Delta k_0 + \left(  v_i^{-1} -  v_p^{-1} \right) \Omega_i + \left(v_p^{-1} - v_o^{-1}  \right) \Omega_o,
\end{equation}
where $\Delta k_0 = k_i(\overline{\omega}_i) - k_p(\overline{\omega}_p) - k_o(\overline{\omega}_o)$. When $\Delta k_0=0$, the orientation of the PMF can be determined by assuming $\Delta k = 0 $ and solving for the output angular frequency $\Omega_o$ as a function of the input angular frequency $\Omega_i$. This gives
\begin{equation}
\Omega_o = \frac{v_i^{-1}-v_p^{-1}}{v_p^{-1}-v_o^{-1}} \Omega_i.
\end{equation}
The angle, $\Theta$, that the PMF makes with respect to the axis corresponding to the input angular frequency is
\begin{equation}
\Theta = \tan^{-1}{\left( \frac{v_i^{-1}-v_p^{-1}}{v_p^{-1}-v_o^{-1}} \right)}.
\end{equation}
Thus, the PMF is parallel to the input angular frequency axis ($\Theta=0$) when the group velocity of the input field is equal to the group velocity of the pump field, $v_i=v_p$. We note that this group velocity requirement is similar to those for broadband optical parametric amplification (OPA) \cite{Manzoni2016} and engineering spectrally pure bi-photon states from spontaneous parametric downconversion \cite{Graffitti2018}. However, for the OPA process, a quasi-monochromatic pump is used to amplify a broadband signal and generate a broadband idler, and, for engineering spectrally pure bi-photon states, the signal and idler group velocities must be matched. In frequency auto-homogenization, a broadband pump field must be group velocity matched with the input field at the center frequency of the target inhomogeneous spectral distribution.

\subsection{Schmidt Decomposition of JCA and Mode Analysis}

Schmidt decomposition of the JCA provides insight into the modal properties and performance of the frequency auto-homogenization device. Using Schmidt decomposition \cite{law_2000, parker_2000}, the JCA can be expressed as a sum of orthogonal modes such that
\begin{equation}
f\left(\omega_{i}, \omega_{o}\right) = \sum_n \sqrt{\kappa_n} g_n(\omega_{i}) h_n^*(\omega_{o}), \label{eq:SDofJCA}
\end{equation}
where $\{g_n\}$ and $\{h^*_n\}$ are sets of orthonormal basis functions for the spectral amplitudes of the input and output Schmidt modes respectively and $\sqrt{\kappa_n}$ is the Schmidt coefficient corresponding to the $n^\textrm{th}$ mode in the summation. The Schmidt coefficients, $\sqrt{\kappa_n}$, satisfy the normalization condition $\sum_n \kappa_n=1$. The Schmidt number defined as
\begin{equation}\label{eq:SchmidtNumber}
K = \frac{1}{\sum_n \kappa_n^2}
\end{equation}
describes the effective number of populated modes \cite{law_2004}. Finally, the purity is defined as the inverse of the Schmidt number, $P = 1/K$. For our application, the JCA should be pure with one dominant Schmidt mode.

To determine the evolution between input and output Schmidt modes, we can define the respective single-photon annihilation operators as
\begin{align}
    \hat{A}_n &:= \int d\omega_i g_n(\omega_i)\hat{a}(\omega_i) \label{eq:inputSchmidtannihilation} \\
    \hat{B}_n &:= \int d\omega_o h_n(\omega_o)\hat{b}(\omega_o).\label{eq:outputSchmidtannihilation}
\end{align}
Substituting Eqs. \ref{eq:SDofJCA}, \ref{eq:inputSchmidtannihilation}, and \ref{eq:outputSchmidtannihilation} into Eq. \ref{eq:TimeIntegratedHamiltonian} yields
\begin{equation}
    \hat{U} = \exp{\left[ \mathcal{C} \sum_n \sqrt{\kappa_n} \hat{A}_n (\omega_i) \hat{B}^\dagger_n (\omega_o) + H.C. \right]}
\end{equation}
which is a product of beamsplitter interactions between each pair of Schmidt modes. Thus, the equations of motion for the Schmidt-mode operators give sinusoidal evolution between the input and output modes:
\begin{align}
    \hat{A}_n(\mathcal{C}) &= \cos{\left( \frac{\mathcal{C}\sqrt{\kappa_n}}{\hbar} \right)} \hat{A}_n(0) + i \sin{\left( \frac{\mathcal{C}\sqrt{\kappa_n}}{\hbar} \right)} \hat{B}_n(0) \\
    \hat{B}_n(\mathcal{C}) &= \cos{\left( \frac{\mathcal{C}\sqrt{\kappa_n}}{\hbar} \right)} \hat{B}_n(0) + i \sin{\left( \frac{\mathcal{C}\sqrt{\kappa_n}}{\hbar} \right)} \hat{A}_n(0).
\end{align}

To find the probability of converting an input photon into the $n^{th}$ Schmidt mode, we calculate $\an{\hat{B}_n^\dagger(\mathcal{C}) \hat{B}_n(\mathcal{C})}$ by first applying $\hat{B}_n(\mathcal{C})$ to the initial state $\ket{\psi_i}$,
\begin{equation}
    \hat{B}_n(\mathcal{C})\ket{\psi_i} = i\sin{\left(  \frac{\mathcal{C}\sqrt{\kappa_n}}{\hbar} \right)} \int d\omega_i \alpha^*(\omega_{i}) g_n(\omega_{i}) \ket{0, 0}.
\end{equation}
Then, the probability of conversion to the $n^{th}$ Schmidt mode is
\begin{equation} \label{eq:SMconversionprobability}
    \eta_n = \an{\hat{B}_n^\dagger (\mathcal{C}) \hat{B}_n(\mathcal{C})} = \sin^2{\left( \frac{\mathcal{C}\sqrt{\kappa_n}}{\hbar} \right)} \left| \int d\omega_i \alpha^*(\omega_i) g_n(\omega_i) \right|^2.
\end{equation}
Finally, the total probability of conversion is the sum of the probabilities into each Schmidt mode,
\begin{equation} \label{eq:totalconversionprobability}
    \eta_\mathrm{tot} = \sum_n \eta_n.
\end{equation}

\subsection{Quantifying Homogenization} \label{sec:quantifyinghomogenization}
The degree of homogenization can be quantified by measuring the interferometric visibility of the cross-correlation between the output field, $E_\mathrm{o,GVM}(t)$, generated from the input field at the group-velocity-matched frequency and all other output fields, $E_\mathrm{o}(t)$. The resulting interferogram $T(\tau)$ is given by
\begin{equation}
    T(\tau)=\left|\int_{-\infty}^{\infty}dt E_\mathrm{o,GVM}(t)E_\mathrm{o}(t+\tau)\right|^2,
\end{equation}
and the visibility is given by
\begin{equation} \label{eq:visibility}
    V=\frac{\mathrm{max}(T)-\mathrm{min}(T)}{\mathrm{max}(T)+\mathrm{min}(T)}.
\end{equation}
In the following sections, this metric is used to quantify the amount of spectral distinguishability that can be removed by the auto-homogenization device.

\section{Calculations and Numerical Simulations}
A 2.5-mm periodically-poled rubidium-doped potassium titanyl phosphate (Rb:KTP) waveguide is studied as a proof-of-principle demonstration. The waveguide is oriented such that propagation is along the crystal X axis with horizontally polarized light along the crystal Y axis and vertically polarized light along the crystal Z axis. The purity of the JCA is calculated using Schmidt decomposition, and the performance is analyzed using a split-step Fourier pulse propagation simulation. Simulations are performed for several narrowband input spectral intensities with different carrier frequencies spaced around the group-velocity-matched input frequency. The pump spectral intensity is kept fixed for all simulations of a single device. The degree of homogenization is evaluated using the interferrometric visibility between output fields as described in Sec. \ref{sec:quantifyinghomogenization}.

\begin{figure*}[t]
\includegraphics[width=\textwidth]{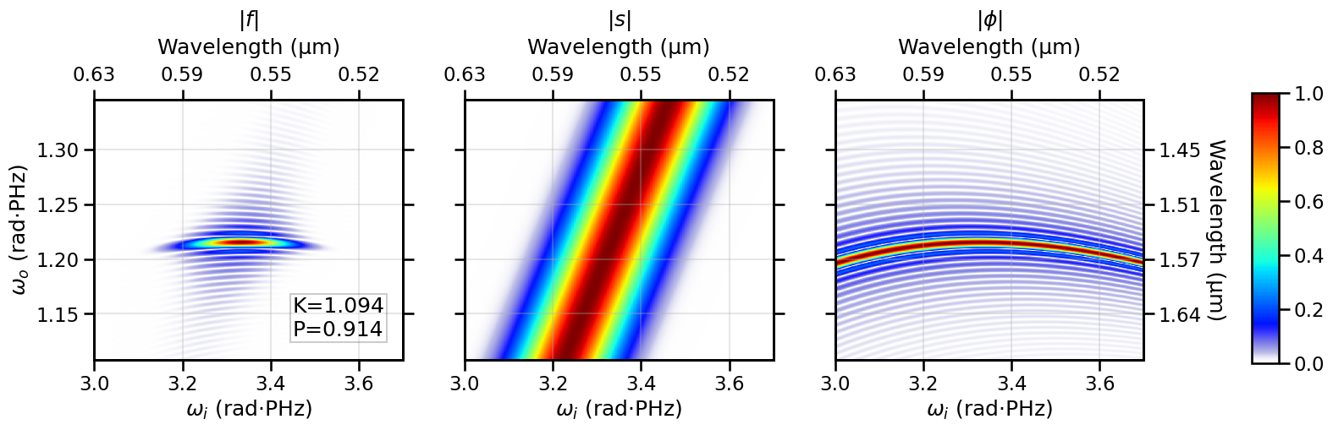}
\caption{Absolute value of the JCA ($f$), spectral pump amplitude ($s$), and the PMF ($\phi$) for a 2.5-mm Rb:KTP waveguide with group-velocity-matched input and pump fields. The pump spectral intensity is assumed to be Gaussian with a 50-nm FWHM bandwidth centered at 0.89 \textmu m. All plots are normalized to the maximum values. The Schmidt number and purity of the coupling amplitude are 1.094 and 0.914 respectively.\label{fig:RbKTPcouplingamp}}
\end{figure*}

For a target output wavelength of 1.55 \textmu m, the input and pump group velocities are matched for $\lambda_i$=0.565 \textmu m and $\lambda_p$=0.89 \textmu m in the Rb:KTP waveguides. The polarizations of the input, output, and pump fields are (o), (o), and (e) respectively. The refractive indices are determined from data provided by AdvR Inc, and the JCA is calculated according to Eq. \ref{eq:JCA} given a Gaussian pump spectral intensity with a 50-nm FWHM bandwidth centered at 0.89 \textmu m. The absolute values of the JCA ($f$), pump spectral amplitude ($s$), and PMF ($\phi$) are shown in Fig. \ref{fig:RbKTPcouplingamp}. The Schmidt number is 1.094, and the purity is 0.914.

\begin{figure}[t]
\includegraphics[width=\textwidth]{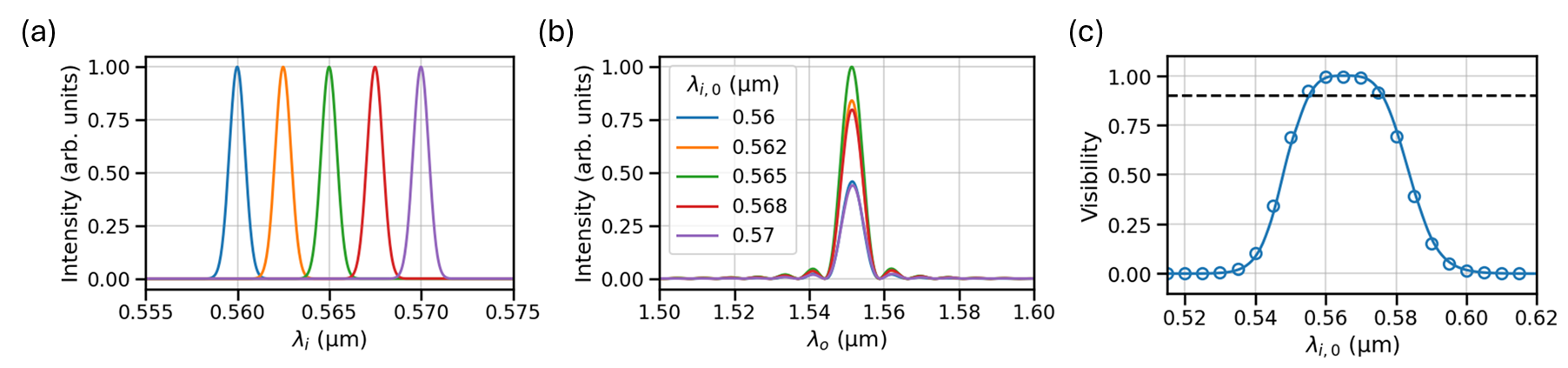}
\caption{(a) Example inhomogeneous input spectra with 1-nm FWHM bandwidths before homogenization and (b) resulting output spectra from split-step Fourier pulse propagation simulations for the 2.5-mm Rb:KTP waveguide after homogenization. The input center wavelength, $\lambda_{i,0}$, is varied about the group-velocity matched wavelength. (c) Visibility of cross-correlations between the output spectra generated from the input pulse with a center wavelength of 0.565 \textmu m and the input pulses with center wavelengths $\lambda_{i,0}$. The solid line shows the visibility calculated from the coupling amplitudes, and the open circles show the visibility calculated from the pulse propagation simulations. The black dashed line marks a visibility of $V=0.9$. For a 50-nm pump, the visibility is above 0.9 over a range of $\sim$20 nm.\label{fig:KTPsimoutput}}
\end{figure}

Split-step Fourier pulse propagation simulations are performed to verify the homogenization capabilities of the device. For all simulations, the pump pulse parameters are kept fixed while the center wavelength of the input pulse, $\lambda_{i,0}$, is varied. The pump pulse is assumed to have a Gaussian profile with a 50-nm FWHM bandwidth centered at 0.89 \textmu m. The pump beam size, intensity, and energy are determined by the mode-field diameter, conversion rate, and crystal length. Several single photon input bandwidths are considered to represent different single photon sources. The input pulse is assumed to have a Gaussian profile with a 0.1-nm, 1-nm, and 5-nm FWHM spectral bandwidth corresponding to transform-limited pulse durations of $\sim$4.7 ps, $\sim$0.47 ps, and $\sim$0.09 ps respectively. For all cases, the input pulse energy is set equal to that of a single photon, $E=\hbar\omega_i$, where $\hbar$ is Planck's constant divided by $2\pi$. The input beam size is assumed to be half that of the pump beam size. 1-nm input spectra and the generated output spectra are shown in Fig. \ref{fig:KTPsimoutput}(a) and \ref{fig:KTPsimoutput}(b) respectively. The line colors in Fig. \ref{fig:KTPsimoutput}(a) correspond to those in \ref{fig:KTPsimoutput}(b).
The interferometric visibility from Eq. \ref{eq:visibility} is used to quantify the degree of homogenization. The visibility between the output field generated by the input field with a center wavelength of $\lambda_{i,0}=0.565$ µm and output fields generated from other input fields is plotted in Fig. \ref{fig:KTPsimoutput}(c) for the case using 1-nm input bandwidths. The visibilities using 0.1-nm and 5-nm input bandwidths are found to be the same as that for 1 nm. In Fig. \ref{fig:KTPsimoutput}(c), the open circles show the results of the pulse propagation simulations assuming a 50-nm FWHM pump spectrum; the solid lines show the visibility calculated using the coupling amplitudes assuming a 50-nm FWHM pump spectrum; and the black dashed line corresponds to a visibility of $V=0.9$. For the 50-nm pump, the visibility calculated using the coupling amplitude lies above this threshold for $\sim$20 nm.

\begin{figure}[t]
\includegraphics[width=7cm]{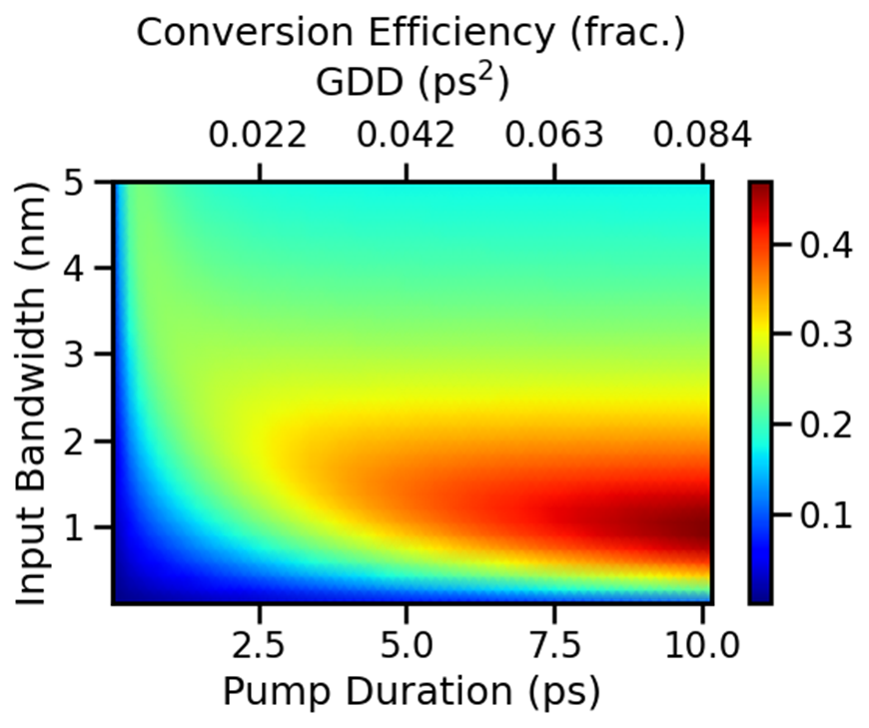}
\caption{Conversion efficiencies from pulse propagation simulations for input fields of increasing bandwidth and pump pulses with increasing normal group delay dispersion (GDD). For all simulations, the pump bandwidth and intensity are kept fixed. \label{fig:conveffsim}}
\end{figure}

Notably, the input bandwidth does not affect the overall homogenization bandwidth as determined by the calculated visibility. However, the input bandwidth does affect the conversion efficiency of the homogenization device. From Eq. \ref{eq:SMconversionprobability}, the conversion probability to each Schmidt mode is proportional to the inner product of the spectral amplitudes of the Schmidt mode and the input field. Considering the JCA shown in Fig. \ref{fig:RbKTPcouplingamp}, the bandwidth of the pump determines the width along the input frequency dimension. In the time domain, this broadband pump has a short TL pulse duration ($\sim$20 fs), and the input fields have varying TL pulse durations ranging from $\sim$0.09 ps to $\sim$4.7 ps for bandwidths ranging from 5 nm to 0.1 nm. Because the pump and input frequencies are group velocity matched, the pump pulse only has the opportunity to convert the input field that lies within its temporal envelope---any probability amplitude outside of the pump's temporal envelope does not have a chance of being converted. This leads to a low conversion efficiency for input fields with bandwidths much smaller than the pump bandwidth. From our simulations, a 1-nm input field is converted to 1550 nm with an efficiency of $\sim0.7\%$ using a TL pump pulse. The efficiency can be increased by increasing the temporal overlap between the input and pump fields. This can be achieved by (1) reducing the pump bandwidth and increasing the pump pulse duration at the expense of the homogenization bandwidth or (2) chirping the pump pulse at the expense of added complexity. If the pump pulse is chirped, each input frequency is efficiently converted when it temporally overlaps with the phase-matched pump frequency. Conversion efficiencies calculated from pulse propagation simulations using input pulses at the group-velocity-matched input frequency with different bandwidths and pump pulses with increasing normal second-order dispersion are plotted in Fig. \ref{fig:conveffsim}. For a 1-nm input bandwidth, a 50-nm pump chirped to 10 ps results in a conversion efficiency of $\sim46.6\%$.

\section{Experiments and Results}

\begin{figure}[t]
\includegraphics[width=\textwidth]{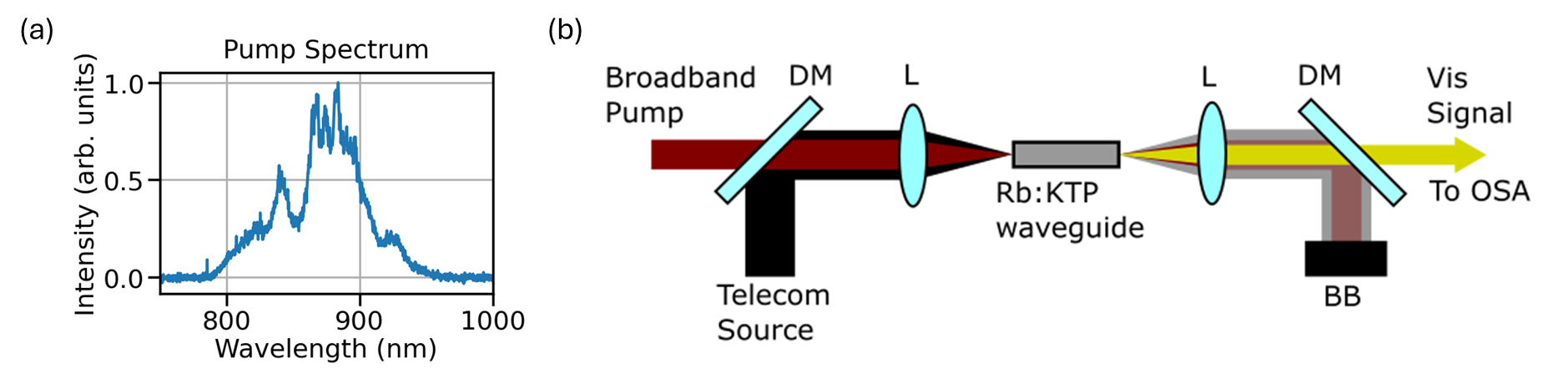}
\caption{Experimental setup showing (a) the broadened pump spectrum used to drive the conversion process and (b) the experimental layout for combining the fields into the Rb:KTP waveguide. DM - Dichroic Mirror, L - Lens, BB - Beam Block. \label{fig:KTPsetup}}
\end{figure}

\begin{figure}[t]
\includegraphics[width=0.8\textwidth]{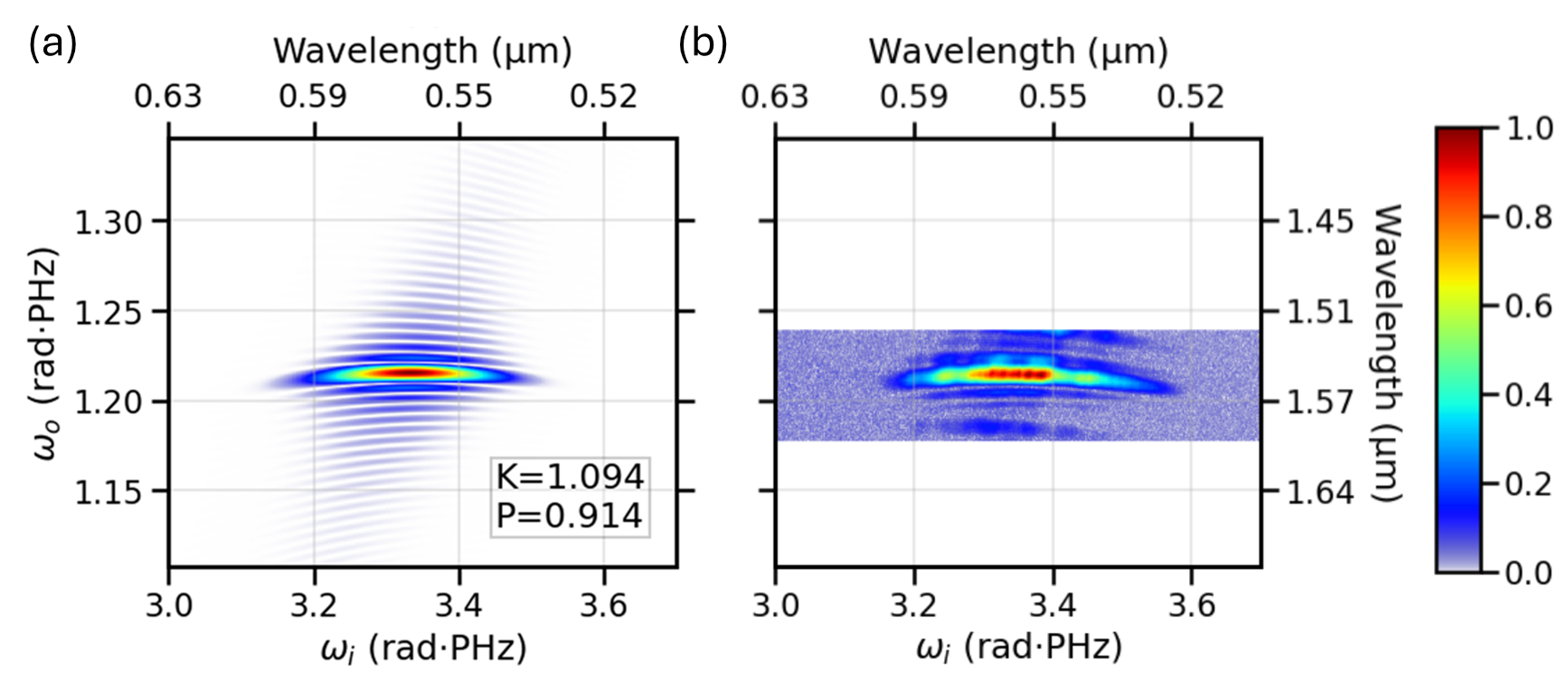}
\caption{Comparison of the magnitudes of (a) the calculated JCA and (b) the measured JCA. \label{fig:JCIcompare}}
\end{figure}

The magnitude of the JCA was measured for 3 frequency homogenization devices, each consisting of 30 separate Rb:KTP waveguides (AdvR Inc). On each device, the waveguides were separated into groups based on the Rb diffusion depth with each group consisting of 6 waveguides: 2 with 2-\textmu m width, 2 with 3-\textmu m width, and 2 with 4-\textmu m width. The experimental setup is shown in Fig. \ref{fig:KTPsetup}.
A wavelength-tunable continuous-wave (CW) laser operating in the telecommunication C band was combined with a broadband pump pulse centered at 875 nm. The CW telecom laser was scanned from 1.52 $\mathrm{\mu}$m to 1.6 $\mathrm{\mu}$m while recording the visible output spectra. All measured JCA magnitudes are compiled in Appendix \ref{JCAmeasurements}, and a representative measurement from one of the waveguides is shown in Fig. \ref{fig:JCIcompare} beside the analytically calculated JCA for comparison. The measured JCA displays the characteristics of the horizontal PMF which is cropped on either side by the bandwidth of the pump pulse. This measurement confirms that the necessary PMF and pump amplitude for homogenization have been achieved. Future experiments include second-order correlation measurements between outputs of multiple homogenization devices.

\section{Discussion}

In this article, we proposed frequency auto-homogenization and presented preliminary measurements of the JCI of Rb:KTP waveguides that confirm the required PMF and pump spectral intensity to convert photons from an inhomogeneous spectral distribution at 565 nm to a single spectral mode at 1550 nm.

Frequency auto-homogenization utilizes group-velocity-matched downconversion with a broadband pump field to convert spectrally inhomogeneous photons to a homogeneous spectral distribution opening a path toward increasing the yield of indistinguishable photons from currenlty available spectrally inhomogeneous single-photon sources. Schmidt decomposition of the JCA and numerical simulations were used to provide insight into the conversion properties and inform device design. High visibility between output modes is achieved when the purity of the JCA is close to unity, corresponding to cases where the input and output fields are uncorrelated. Additionally, a broadband pump is required to achieve homogenization over the broadest possible bandwidth by satisfying energy conservation between all possible input frequencies and the target output frequency. Thus, the effective bandwidth of the device is determined by the flatness of the PMF (i.e. the difference in group delay between the output frequency and the group-velocity-matched pump and input frequencies) and the bandwidth of the broadband pump. From our simulations, we find that photons from a 10-nm (20-nm) inhomogeneous spectral distribution centered at 565 nm can be converted to output spectra at 1550 nm with visibilities greater than $99\%$ ($90\%$).

The conversion efficiency of the device is inherently limited by the overlap between the input Schmidt mode of the JCA and the input photon spectrum. Never-the-less, we note that the efficiency can be increased by either (1) reducing the pump bandwidth at the cost of decreasing the homogenization bandwidth, or (2) chirping the pump pulse at the cost of increased complexity and temporal distinguishability. If the pump pulse is chirped, the individual input frequencies must be delayed to coincide with their respective pump frequencies as dictated by energy conservation, requiring a priori information and yielding temporally distinguishable outputs. Moreover, we anticipate that the efficiency and homogenization bandwidth can be optimized for the desired application on a case-by-case basis and expect that pump spectral and pulse shaping can be used to further increase device performance. Additionally, we note that birefringent phase matching in bulk nonlinear crystals may be used to achieve auto-frequency homogenization for other input wavelengths and have compiled a list of crystals and the required PM angles in Table \ref{tab:horizontalPMF} in Appendix \ref{CrystalTable}. Finally, dispersion engineering in integrated photonic platforms may also be used to achieve homogenization in other wavelength ranges. If periodic poling is used to achieve phase matching, the strong pump field can generate a broad spontaneous parametric downconversion spectrum at lower frequencies due to variations in fabrication that may overlap with the target output spectrum \cite{pelc2011long}. This added noise should be considered during device design and is avoided if the pump wavelength is longer than the output wavelength. Researchers have addressed noise in single-photon downconversion using cascaded long-wavelength pumping \cite{esfandyarpour2018cascaded} or bulk devices with resonant enhancement \cite{geus2024low}.

We emphasize that this conversion method homogenizes the spectra of quantum emitters erasing their inherent spectral distinguishability without a priori knowledge of their carrier frequency within the homogenization bandwidth nor the need for individualized tuning. By engineering the JCA of the downconversion process, frequency auto-homogenization constitutes a means of increasing the yield of indistinguishable photons and scalability of current state-of-the-art integrated sources for diverse applications in quantum information science.

\appendix
\section{Interaction Hamiltonian and Joint Coupling Amplitude} \label{app:HintJCA}

For frequency auto-homogenization, DFG is used to downconvert a high-frequency input photon from an inhomogeneously broadened spectral distribution to a low-frequency output photon having a homogeneous spectral distribution using a broadband, classically bright pump field. The angular frequencies of these fields are given by $\omega_i$, $\omega_o$, and $\omega_p$ respectively. This process is described by the unitary time-evolution operator for quantum frequency downconversion
\begin{equation} \label{eq:TEOAppen1}
    \hat{U}(0,t)= \exp{\left(-\frac{i}{\hbar}\int_0^t{dt'\hat{H}_{FC}}\right)}.
\end{equation}
Following \cite{Schneeloch_2019}, the frequency conversion Hamiltonian is given by
\begin{align} \label{eq:HFCAppen1}
    \hat{H}_{FC}=i\hbar\sqrt{\frac{\hbar d_{\mathrm{eff}}^2}{4\pi^3\epsilon_0 c^3}}\int  & d\omega_i d\omega_o d\omega_p \sqrt{\frac{\omega_i \omega_o \omega_p  n_{g,i} n_{g,o} n_{g,p} }{ n_i^2 n_o^2 n_p^2 }} \nonumber \\ 
    & \times S^*(\omega_p) \Phi(\Delta k) \exp{\left(-i \Delta\omega t \right)} \hat{a}(\omega_i)\hat{b}^\dagger(\omega_o) + H.C.
\end{align}
where $d_\mathrm{eff}$ is the effective nonlinear coefficient of the medium, $n_j$ and $n_{g,j}$ for $j=i,o,p$ are the refractive index and group index of each field respectively, $S^*(\omega_p)$ is the conjugated spectral amplitude of the classically bright pump field, $\Phi(\Delta k)$ is the phase-matching function, $\hat{a}(\omega_i)$ and $\hat{b}^\dagger(\omega_o)$ are the respective annihilation and creation operators for photons in the input and output modes at their respective frequencies, and $H.C.$ is the Hermitian conjugate. Here, we assume that the angular frequencies, refractive indices, and group indices vary slowly with respect to the other functions in the integrand and can therefore be evaluated at their center angular frequencies, $\overline{\omega}_j$, and taken out of the integral giving
\begin{align} \label{eq:HFCAppen2}
    \hat{H}_{FC} \approx i\hbar&\sqrt{\frac{\hbar d_{\mathrm{eff}}^2}{4\pi^3\epsilon_0 c^3}}  \sqrt{\frac{\overline{\omega}_i \overline{\omega}_o \overline{\omega}_p  n_{g,i} n_{g,o} n_{g,p} }{ n_i^2 n_o^2 n_p^2 }} \nonumber \\
    & \times \int d\omega_i d\omega_o d\omega_p S^*(\omega_p) \Phi(\Delta k) \exp{\left(-i \Delta\omega t \right)} \hat{a}(\omega_i)\hat{b}^\dagger(\omega_o) + H.C.
\end{align}

To ensure that the joint coupling amplitude is normalized, the pump spectral amplitude and phase-matching function must be considered in more detail. First, the pump spectral amplitude $S(\omega_p)$ is normalized such that 
\begin{equation} \label{eq:PSAAppen1}
    S{(\omega_p)}=\sqrt{N_p}s(\omega_p),
\end{equation}
where $N_p$ is the photon number in the pump field and $s(\omega_p)$ is the normalized pump spectral amplitude such that $\int d\omega_p|s(\omega_p)|^2=1$. Second, the phase-matching function is given by
\begin{equation} \label{eq:PMFAppen1}
    \Phi\left(\Delta k\right) \equiv \int d^3r\left[g_{i}(x,y) g_{o}^*(x,y)g_{p}^*(x,y)\exp{\left( i \Delta k z \right)} \right],
\end{equation}
where $g_j(x,y)$ is the transverse mode profile of field $j$ and $\Delta k$ is the wave-vector mismatch along the propagation direction. By assuming collimated beams with Gaussian profiles and that the interaction takes place well within the Rayleigh range of the smallest beam, Eq. \ref{eq:PMFAppen1} can be greatly simplified:
\begin{align} \label{eq:PMFAppen2}
    \Phi(\Delta k) &= \sqrt{\frac{2}{\pi}} \frac{\sigma_i \sigma_o \sigma_p}{\sigma_i^2\sigma_o^2+\sigma_i^2\sigma_p^2+\sigma_o^2\sigma_p^2} \int_{-L/2}^{L/2} dz \exp{\left(i\Delta k z\right)} \nonumber \\
    &= \sqrt{\frac{2}{\pi}} \frac{\sigma_i \sigma_o \sigma_p L}{\sigma_i^2\sigma_o^2+\sigma_i^2\sigma_p^2+\sigma_o^2\sigma_p^2}  \operatorname{sinc}{\left(\frac{\Delta k L}{2}\right)} \\
     &= \sqrt{\frac{2}{\pi}} \frac{\sigma_i \sigma_o \sigma_p L \mathcal{N}}{\sigma_i^2\sigma_o^2+\sigma_i^2\sigma_p^2+\sigma_o^2\sigma_p^2}  \phi{(\Delta k)},
\end{align}
where $\sigma_j$ is the standard-deviation beam radius of beam $j$, $L$ is the length of the nonlinear medium, and $\phi(\Delta k)=\operatorname{sinc}{\left(\frac{\Delta k L} {2}\right)} / \mathcal{N}$ is the normalized PMF.

Substituting Eqs. \ref{eq:PSAAppen1} and \ref{eq:PMFAppen2} into Eq. \ref{eq:HFCAppen2} and performing the integration over time in Eq. \ref{eq:TEOAppen1}, noting that the interaction is zero outside of the integration limits, yields
\begin{equation} \label{eq:TEOAppen2}
    \hat{U} \approx \exp{\left( \mathcal{C} \int d\omega_i d\omega_o s^*(\omega_i-\omega_o) \phi(\omega_i,\omega_o) \hat{a}(\omega_i) \hat{b}^\dagger(\omega_o) + H.C. \right)}
\end{equation}
where
\begin{equation}
    \mathcal{C} = \sqrt{\frac{2\hbar d_\mathrm{eff}^2}{\pi^2\epsilon_0 c^3}} \sqrt{\frac{\overline{\omega}_i \overline{\omega}_o \overline{\omega}_p  n_{g,i} n_{g,o} n_{g,p} }{ n_i^2 n_o^2 n_p^2 }} \frac{\sigma_i \sigma_o \sigma_p L \mathcal{N} \sqrt{N_p}}{\sigma_i^2\sigma_o^2+\sigma_i^2\sigma_p^2+\sigma_o^2\sigma_p^2}.
\end{equation}
Note, in the derivation of Eq. \ref{eq:TEOAppen2}, we invoke the delta function $\int_{-\infty}^{\infty}dt\exp{\left(-i \Delta\omega t\right)} = 2\pi\delta(\Delta\omega)$ and explicitly show the frequency dependence of the PMF. Finally from Eq. \ref{eq:TEOAppen2}, the JCA is defined as 
\begin{equation}
    f(\omega_i, \omega_o) = s^*(\omega_i - \omega_o) \phi(\omega_i, \omega_o).
\end{equation}

\section{Homogenization with Birefringent Phase-Matching} \label{CrystalTable}
The group velocities of the pump and input fields are equal for the crystals, wavelengths, and phase-matching conditions shown in Table \ref{tab:horizontalPMF}. Note, this list is not exhaustive and only considers bulk, unpoled crystals with an output wavelength of 1.55 \textmu m. Waveguide dispersion and quasi-phase matching may allow the group-velocity matching condition to be met for other materials and wavelengths.
\begin{table*}[ht]
\begin{ruledtabular}
\begin{tabular}{|c|c|c|c|c|c|c|c|}
\hline
Crystal & $\lambda_i$ (nm) & $\lambda_o$ (nm) & $\lambda_p$ (nm) & $\theta$ ($^\circ$) & $\phi$ ($^\circ$) & PP & $n$ Ref. \\
\hline{}
BBO & 908 (e) & 1550 (o) & 2192.2 (e) & 26.7 & — & — & \cite{tamosauskas_2018} \\
\hline{}
BBO & 902 (e) & 1550 (o) & 2157.6 (e) & 27.4 & — & — & \cite{kato_1986} \\
\hline
BiBO & 705 (o) & 1550 (o) & 1293.2 (e) & 44 & 0 & XZ & \cite{umemura_2007} \\
\hline
BiBO & 969 (o) & 1550 (e) & 2585.1 (e) & 41.3 & 0 & XZ & \cite{umemura_2007} \\
\hline
BiBO & 885 (o) & 1550 (e) & 2062.8 (e) & 34.9 & — & XZ & \cite{umemura_2007} \\
\hline
LBO & 640 (o) & 1550 (e) & 1090.1 (e) & 87.7 & 0 & XZ & \cite{kato_1994} \\
\hline
LBO & 520 (e) & 1550 (o) & 782.5 (o) & 22.3 & 0 & XZ & \cite{kato_1994} \\
\hline
LBO & 840 (o) & 1550 (e) & 1833.8 (o) & 15.4 & 90 & YZ & \cite{kato_1994} \\
\hline
LiIO$_3$ & 815 (e) & 1550 (e) & 1718.7 (o) & 30.6 & — & — & \cite{kato_1985} \\
\hline
KTP & 745 (o) & 1550 (o) & 1434.5 (e) & 53 & 0 & XZ & \cite{kato_2002} \\
\hline
KTP & 735 (o) & 1550 (o) & 1397.9 (e) & 44 & 90 & YZ & \cite{kato_2002} \\
\hline
Mg:LNB & 961 (e) & 1550 (o) & 2528.9 (o) & 43 & — & — & \cite{zelmon_1997} \\
\hline
\end{tabular}
\end{ruledtabular}
\caption{Crystals, wavelengths, and phase-matching parameters that result in group-velocity matching between the input and pump fields and an output wavelength of 1.55 \textmu m. The input, output, and pump wavelengths are $\lambda_i$, $\lambda_o$, and $\lambda_p$, respectively. The polarization of each field is listed as either e for extraordinary or o for ordinary, where the extraordinary polarization lies in the plane created by the wave vector and the crystal axis for uniaxial crystals or in the principal plane (PP) for biaxial crystals. For uniaxial crystals, the phase-matching angle $\theta$ defines the angle between the crystal axis and the wave vector. For biaxial crystals, the phase-matching angles $\theta$ and $\phi$ define the direction of the wave vector from the crystal Z axis to the XY plane and from the crystal X axis to the crystal Y axis, where the crystal axes are defined such that $n_\mathrm{X}<n_\mathrm{Y}<n_\mathrm{Z}$. The phase-matching angle $\phi$ and the principal plane are only listed for biaxial crystals.}
\label{tab:horizontalPMF}
\end{table*}

\section{Compilation of JCA measurements} \label{JCAmeasurements}

\begin{figure*}[htbp]
\includegraphics[width=0.95\textwidth]{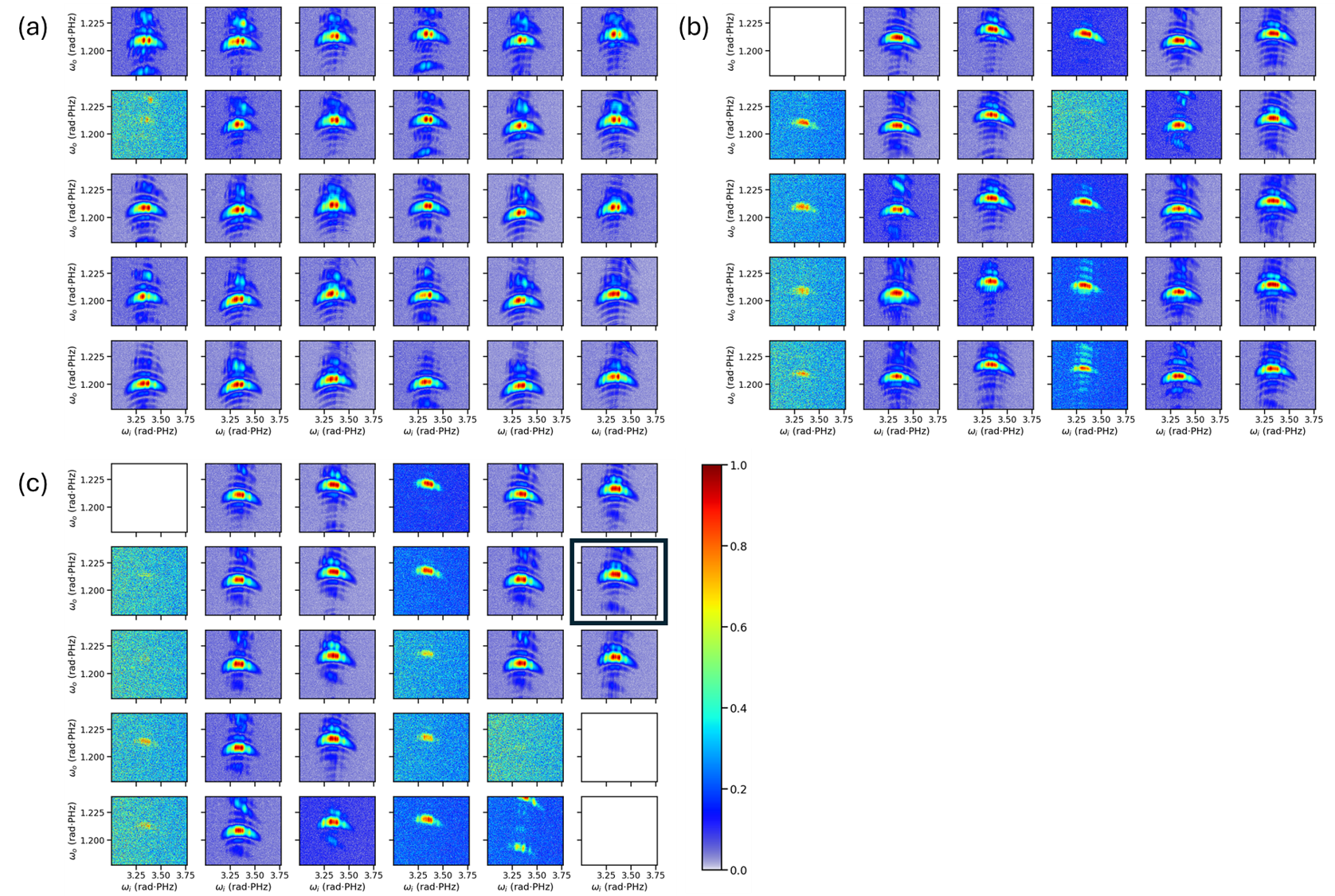}
\caption{JCA magnitudes measured for (a) device 1, (b) device 2, and (c) device 3. The black outlined plot in (c) shows the image presented in the main text.}
\label{fig:AllJCA}
\end{figure*}

The magnitude of the JCA was measured for 3 frequency homogenization devices, each consisting of 30 separate Rb:KTP waveguides (AdvR Inc). The JCA magnitudes for each device normalized to their maximum values are plotted in Fig. \ref{fig:AllJCA}. The black outline in Fig. \ref{fig:AllJCA}(c) shows the plot presented in the main text in Fig. \ref{fig:JCIcompare}. Empty plots are cases where generation of visible light was not observed.

\bibliography{main}

\providecommand{\noopsort}[1]{}\providecommand{\singleletter}[1]{#1}%
\begin{thebibliography}{60}%
\makeatletter
\providecommand \@ifxundefined [1]{%
 \@ifx{#1\undefined}
}%
\providecommand \@ifnum [1]{%
 \ifnum #1\expandafter \@firstoftwo
 \else \expandafter \@secondoftwo
 \fi
}%
\providecommand \@ifx [1]{%
 \ifx #1\expandafter \@firstoftwo
 \else \expandafter \@secondoftwo
 \fi
}%
\providecommand \natexlab [1]{#1}%
\providecommand \enquote  [1]{``#1''}%
\providecommand \bibnamefont  [1]{#1}%
\providecommand \bibfnamefont [1]{#1}%
\providecommand \citenamefont [1]{#1}%
\providecommand \href@noop [0]{\@secondoftwo}%
\providecommand \href [0]{\begingroup \@sanitize@url \@href}%
\providecommand \@href[1]{\@@startlink{#1}\@@href}%
\providecommand \@@href[1]{\endgroup#1\@@endlink}%
\providecommand \@sanitize@url [0]{\catcode `\\12\catcode `\$12\catcode `\&12\catcode `\#12\catcode `\^12\catcode `\_12\catcode `\%12\relax}%
\providecommand \@@startlink[1]{}%
\providecommand \@@endlink[0]{}%
\providecommand \url  [0]{\begingroup\@sanitize@url \@url }%
\providecommand \@url [1]{\endgroup\@href {#1}{\urlprefix }}%
\providecommand \urlprefix  [0]{URL }%
\providecommand \Eprint [0]{\href }%
\providecommand \doibase [0]{https://doi.org/}%
\providecommand \selectlanguage [0]{\@gobble}%
\providecommand \bibinfo  [0]{\@secondoftwo}%
\providecommand \bibfield  [0]{\@secondoftwo}%
\providecommand \translation [1]{[#1]}%
\providecommand \BibitemOpen [0]{}%
\providecommand \bibitemStop [0]{}%
\providecommand \bibitemNoStop [0]{.\EOS\space}%
\providecommand \EOS [0]{\spacefactor3000\relax}%
\providecommand \BibitemShut  [1]{\csname bibitem#1\endcsname}%
\let\auto@bib@innerbib\@empty
\bibitem [{\citenamefont {Ralph}\ \emph {et~al.}(2001)\citenamefont {Ralph}, \citenamefont {White}, \citenamefont {Munro},\ and\ \citenamefont {Milburn}}]{ralph_2001}%
  \BibitemOpen
  \bibfield  {author} {\bibinfo {author} {\bibfnamefont {T.~C.}\ \bibnamefont {Ralph}}, \bibinfo {author} {\bibfnamefont {A.~G.}\ \bibnamefont {White}}, \bibinfo {author} {\bibfnamefont {W.~J.}\ \bibnamefont {Munro}},\ and\ \bibinfo {author} {\bibfnamefont {G.~J.}\ \bibnamefont {Milburn}},\ }\bibfield  {title} {\bibinfo {title} {Simple scheme for efficient linear optics quantum gates},\ }\href {https://doi.org/10.1103/PhysRevA.65.012314} {\bibfield  {journal} {\bibinfo  {journal} {Phys. Rev. A}\ }\textbf {\bibinfo {volume} {65}},\ \bibinfo {pages} {012314} (\bibinfo {year} {2001})}\BibitemShut {NoStop}%
\bibitem [{\citenamefont {Browne}\ and\ \citenamefont {Rudolph}(2005)}]{browne_2005}%
  \BibitemOpen
  \bibfield  {author} {\bibinfo {author} {\bibfnamefont {D.~E.}\ \bibnamefont {Browne}}\ and\ \bibinfo {author} {\bibfnamefont {T.}~\bibnamefont {Rudolph}},\ }\bibfield  {title} {\bibinfo {title} {Resource-efficient linear optical quantum computation},\ }\href {https://doi.org/10.1103/PhysRevLett.95.010501} {\bibfield  {journal} {\bibinfo  {journal} {Phys. Rev. Lett.}\ }\textbf {\bibinfo {volume} {95}},\ \bibinfo {pages} {010501} (\bibinfo {year} {2005})}\BibitemShut {NoStop}%
\bibitem [{\citenamefont {Knill}(2004)}]{knill_2004}%
  \BibitemOpen
  \bibfield  {author} {\bibinfo {author} {\bibfnamefont {E.}~\bibnamefont {Knill}},\ }\href {https://doi.org/10.48550/ARXIV.QUANT-PH/0404104} {\bibinfo {title} {Fault-tolerant postselected quantum computation: Threshold analysis}} (\bibinfo {year} {2004})\BibitemShut {NoStop}%
\bibitem [{\citenamefont {Rohde}\ and\ \citenamefont {Ralph}(2005)}]{rohde_2005}%
  \BibitemOpen
  \bibfield  {author} {\bibinfo {author} {\bibfnamefont {P.~P.}\ \bibnamefont {Rohde}}\ and\ \bibinfo {author} {\bibfnamefont {T.~C.}\ \bibnamefont {Ralph}},\ }\bibfield  {title} {\bibinfo {title} {Frequency and temporal effects in linear optical quantum computing},\ }\href {https://doi.org/10.1103/PhysRevA.71.032320} {\bibfield  {journal} {\bibinfo  {journal} {Phys. Rev. A}\ }\textbf {\bibinfo {volume} {71}},\ \bibinfo {pages} {032320} (\bibinfo {year} {2005})}\BibitemShut {NoStop}%
\bibitem [{\citenamefont {Dawson}\ \emph {et~al.}(2006)\citenamefont {Dawson}, \citenamefont {Haselgrove},\ and\ \citenamefont {Nielsen}}]{dawson_2006}%
  \BibitemOpen
  \bibfield  {author} {\bibinfo {author} {\bibfnamefont {C.~M.}\ \bibnamefont {Dawson}}, \bibinfo {author} {\bibfnamefont {H.~L.}\ \bibnamefont {Haselgrove}},\ and\ \bibinfo {author} {\bibfnamefont {M.~A.}\ \bibnamefont {Nielsen}},\ }\bibfield  {title} {\bibinfo {title} {Noise thresholds for optical quantum computers},\ }\href {https://doi.org/10.1103/PhysRevLett.96.020501} {\bibfield  {journal} {\bibinfo  {journal} {Phys. Rev. Lett.}\ }\textbf {\bibinfo {volume} {96}},\ \bibinfo {pages} {020501} (\bibinfo {year} {2006})}\BibitemShut {NoStop}%
\bibitem [{\citenamefont {Rohde}\ and\ \citenamefont {Ralph}(2006)}]{rohde_2006}%
  \BibitemOpen
  \bibfield  {author} {\bibinfo {author} {\bibfnamefont {P.~P.}\ \bibnamefont {Rohde}}\ and\ \bibinfo {author} {\bibfnamefont {T.~C.}\ \bibnamefont {Ralph}},\ }\bibfield  {title} {\bibinfo {title} {Error models for mode mismatch in linear optics quantum computing},\ }\href {https://doi.org/10.1103/PhysRevA.73.062312} {\bibfield  {journal} {\bibinfo  {journal} {Phys. Rev. A}\ }\textbf {\bibinfo {volume} {73}},\ \bibinfo {pages} {062312} (\bibinfo {year} {2006})}\BibitemShut {NoStop}%
\bibitem [{\citenamefont {Shchesnovich}(2014)}]{shchesnovich_2014}%
  \BibitemOpen
  \bibfield  {author} {\bibinfo {author} {\bibfnamefont {V.~S.}\ \bibnamefont {Shchesnovich}},\ }\bibfield  {title} {\bibinfo {title} {Sufficient condition for the mode mismatch of single photons for scalability of the boson-sampling computer},\ }\href {https://doi.org/10.1103/PhysRevA.89.022333} {\bibfield  {journal} {\bibinfo  {journal} {Phys. Rev. A}\ }\textbf {\bibinfo {volume} {89}},\ \bibinfo {pages} {022333} (\bibinfo {year} {2014})}\BibitemShut {NoStop}%
\bibitem [{\citenamefont {Shchesnovich}(2015)}]{shchesnovich_2015}%
  \BibitemOpen
  \bibfield  {author} {\bibinfo {author} {\bibfnamefont {V.~S.}\ \bibnamefont {Shchesnovich}},\ }\bibfield  {title} {\bibinfo {title} {Partial indistinguishability theory for multiphoton experiments in multiport devices},\ }\href {https://doi.org/10.1103/PhysRevA.91.013844} {\bibfield  {journal} {\bibinfo  {journal} {Phys. Rev. A}\ }\textbf {\bibinfo {volume} {91}},\ \bibinfo {pages} {013844} (\bibinfo {year} {2015})}\BibitemShut {NoStop}%
\bibitem [{\citenamefont {Tichy}(2014)}]{tichy_2014}%
  \BibitemOpen
  \bibfield  {author} {\bibinfo {author} {\bibfnamefont {M.~C.}\ \bibnamefont {Tichy}},\ }\bibfield  {title} {\bibinfo {title} {Interference of identical particles from entanglement to boson-sampling},\ }\href@noop {} {\bibfield  {journal} {\bibinfo  {journal} {J. Phys. B: At. Mol. Opt. Phys.}\ }\textbf {\bibinfo {volume} {47}},\ \bibinfo {pages} {103001} (\bibinfo {year} {2014})}\BibitemShut {NoStop}%
\bibitem [{\citenamefont {Tichy}(2015)}]{tichy_2015}%
  \BibitemOpen
  \bibfield  {author} {\bibinfo {author} {\bibfnamefont {M.~C.}\ \bibnamefont {Tichy}},\ }\bibfield  {title} {\bibinfo {title} {Sampling of partially distinguishable bosons and the relation to the multidimensional permanent},\ }\href {https://doi.org/10.1103/PhysRevA.91.022316} {\bibfield  {journal} {\bibinfo  {journal} {Phys. Rev. A}\ }\textbf {\bibinfo {volume} {91}},\ \bibinfo {pages} {022316} (\bibinfo {year} {2015})}\BibitemShut {NoStop}%
\bibitem [{\citenamefont {Rohde}(2015)}]{rohde_2015}%
  \BibitemOpen
  \bibfield  {author} {\bibinfo {author} {\bibfnamefont {P.~P.}\ \bibnamefont {Rohde}},\ }\bibfield  {title} {\bibinfo {title} {Boson sampling with photons of arbitrary spectral structure},\ }\href {https://doi.org/10.1103/PhysRevA.91.012307} {\bibfield  {journal} {\bibinfo  {journal} {Phys. Rev. A}\ }\textbf {\bibinfo {volume} {91}},\ \bibinfo {pages} {012307} (\bibinfo {year} {2015})}\BibitemShut {NoStop}%
\bibitem [{\citenamefont {Shi}\ and\ \citenamefont {Byrnes}(2022)}]{shi_2022}%
  \BibitemOpen
  \bibfield  {author} {\bibinfo {author} {\bibfnamefont {J.}~\bibnamefont {Shi}}\ and\ \bibinfo {author} {\bibfnamefont {T.}~\bibnamefont {Byrnes}},\ }\bibfield  {title} {\bibinfo {title} {Effect of partial distinguishability on quantum supremacy in gaussian boson sampling},\ }\bibfield  {journal} {\bibinfo  {journal} {NPJ Quantum Inf.}\ }\textbf {\bibinfo {volume} {8}},\ \href {https://doi.org/10.1038/s41534-022-00557-9} {10.1038/s41534-022-00557-9} (\bibinfo {year} {2022})\BibitemShut {NoStop}%
\bibitem [{\citenamefont {Zhong}\ \emph {et~al.}(2020)\citenamefont {Zhong}, \citenamefont {Wang}, \citenamefont {Deng}, \citenamefont {Chen}, \citenamefont {Peng}, \citenamefont {Luo}, \citenamefont {Qin}, \citenamefont {Wu}, \citenamefont {Ding}, \citenamefont {Hu}, \citenamefont {Hu}, \citenamefont {Yang}, \citenamefont {Zhang}, \citenamefont {Li}, \citenamefont {Li}, \citenamefont {Jiang}, \citenamefont {Gan}, \citenamefont {Yang}, \citenamefont {You}, \citenamefont {Wang}, \citenamefont {Li}, \citenamefont {Liu}, \citenamefont {Lu},\ and\ \citenamefont {Pan}}]{zhong_2020}%
  \BibitemOpen
  \bibfield  {author} {\bibinfo {author} {\bibfnamefont {H.-S.}\ \bibnamefont {Zhong}}, \bibinfo {author} {\bibfnamefont {H.}~\bibnamefont {Wang}}, \bibinfo {author} {\bibfnamefont {Y.-H.}\ \bibnamefont {Deng}}, \bibinfo {author} {\bibfnamefont {M.-C.}\ \bibnamefont {Chen}}, \bibinfo {author} {\bibfnamefont {L.-C.}\ \bibnamefont {Peng}}, \bibinfo {author} {\bibfnamefont {Y.-H.}\ \bibnamefont {Luo}}, \bibinfo {author} {\bibfnamefont {J.}~\bibnamefont {Qin}}, \bibinfo {author} {\bibfnamefont {D.}~\bibnamefont {Wu}}, \bibinfo {author} {\bibfnamefont {X.}~\bibnamefont {Ding}}, \bibinfo {author} {\bibfnamefont {Y.}~\bibnamefont {Hu}}, \bibinfo {author} {\bibfnamefont {P.}~\bibnamefont {Hu}}, \bibinfo {author} {\bibfnamefont {X.-Y.}\ \bibnamefont {Yang}}, \bibinfo {author} {\bibfnamefont {W.-J.}\ \bibnamefont {Zhang}}, \bibinfo {author} {\bibfnamefont {H.}~\bibnamefont {Li}}, \bibinfo {author} {\bibfnamefont {Y.}~\bibnamefont {Li}}, \bibinfo {author} {\bibfnamefont {X.}~\bibnamefont {Jiang}}, \bibinfo {author}
  {\bibfnamefont {L.}~\bibnamefont {Gan}}, \bibinfo {author} {\bibfnamefont {G.}~\bibnamefont {Yang}}, \bibinfo {author} {\bibfnamefont {L.}~\bibnamefont {You}}, \bibinfo {author} {\bibfnamefont {Z.}~\bibnamefont {Wang}}, \bibinfo {author} {\bibfnamefont {L.}~\bibnamefont {Li}}, \bibinfo {author} {\bibfnamefont {N.-L.}\ \bibnamefont {Liu}}, \bibinfo {author} {\bibfnamefont {C.-Y.}\ \bibnamefont {Lu}},\ and\ \bibinfo {author} {\bibfnamefont {J.-W.}\ \bibnamefont {Pan}},\ }\bibfield  {title} {\bibinfo {title} {Quantum computational advantage using photons},\ }\href {https://doi.org/10.1126/science.abe8770} {\bibfield  {journal} {\bibinfo  {journal} {Science}\ }\textbf {\bibinfo {volume} {370}},\ \bibinfo {pages} {1460} (\bibinfo {year} {2020})},\ \Eprint {https://arxiv.org/abs/https://www.science.org/doi/pdf/10.1126/science.abe8770} {https://www.science.org/doi/pdf/10.1126/science.abe8770} \BibitemShut {NoStop}%
\bibitem [{\citenamefont {Wehner}\ \emph {et~al.}(2018)\citenamefont {Wehner}, \citenamefont {Elkouss},\ and\ \citenamefont {Hanson}}]{wehner_2018}%
  \BibitemOpen
  \bibfield  {author} {\bibinfo {author} {\bibfnamefont {S.}~\bibnamefont {Wehner}}, \bibinfo {author} {\bibfnamefont {D.}~\bibnamefont {Elkouss}},\ and\ \bibinfo {author} {\bibfnamefont {R.}~\bibnamefont {Hanson}},\ }\bibfield  {title} {\bibinfo {title} {Quantum internet: A vision for the road ahead},\ }\href {https://doi.org/10.1126/science.aam9288} {\bibfield  {journal} {\bibinfo  {journal} {Science}\ }\textbf {\bibinfo {volume} {362}},\ \bibinfo {pages} {eaam9288} (\bibinfo {year} {2018})},\ \Eprint {https://arxiv.org/abs/https://www.science.org/doi/pdf/10.1126/science.aam9288} {https://www.science.org/doi/pdf/10.1126/science.aam9288} \BibitemShut {NoStop}%
\bibitem [{\citenamefont {Aharonovich}\ \emph {et~al.}(2016)\citenamefont {Aharonovich}, \citenamefont {Englund},\ and\ \citenamefont {Toth}}]{aharonovich_2016}%
  \BibitemOpen
  \bibfield  {author} {\bibinfo {author} {\bibfnamefont {I.}~\bibnamefont {Aharonovich}}, \bibinfo {author} {\bibfnamefont {D.}~\bibnamefont {Englund}},\ and\ \bibinfo {author} {\bibfnamefont {M.}~\bibnamefont {Toth}},\ }\bibfield  {title} {\bibinfo {title} {Solid-state single-photon emitters},\ }\href {https://doi.org/10.1038/nphoton.2016.186} {\bibfield  {journal} {\bibinfo  {journal} {Nature Photonics}\ }\textbf {\bibinfo {volume} {10}},\ \bibinfo {pages} {631} (\bibinfo {year} {2016})}\BibitemShut {NoStop}%
\bibitem [{\citenamefont {Ding}\ \emph {et~al.}(2016)\citenamefont {Ding}, \citenamefont {He}, \citenamefont {Duan}, \citenamefont {Gregersen}, \citenamefont {Chen}, \citenamefont {Unsleber}, \citenamefont {Maier}, \citenamefont {Schneider}, \citenamefont {Kamp}, \citenamefont {Höfling}, \citenamefont {Lu}, ,\ and\ \citenamefont {Pan}}]{Ding_2016}%
  \BibitemOpen
  \bibfield  {author} {\bibinfo {author} {\bibfnamefont {X.}~\bibnamefont {Ding}}, \bibinfo {author} {\bibfnamefont {Y.}~\bibnamefont {He}}, \bibinfo {author} {\bibfnamefont {Z.-C.}\ \bibnamefont {Duan}}, \bibinfo {author} {\bibfnamefont {N.}~\bibnamefont {Gregersen}}, \bibinfo {author} {\bibfnamefont {M.-C.}\ \bibnamefont {Chen}}, \bibinfo {author} {\bibfnamefont {S.}~\bibnamefont {Unsleber}}, \bibinfo {author} {\bibfnamefont {S.}~\bibnamefont {Maier}}, \bibinfo {author} {\bibfnamefont {C.}~\bibnamefont {Schneider}}, \bibinfo {author} {\bibfnamefont {M.}~\bibnamefont {Kamp}}, \bibinfo {author} {\bibfnamefont {S.}~\bibnamefont {Höfling}}, \bibinfo {author} {\bibfnamefont {C.-Y.}\ \bibnamefont {Lu}}, ,\ and\ \bibinfo {author} {\bibfnamefont {J.-W.}\ \bibnamefont {Pan}},\ }\bibfield  {title} {\bibinfo {title} {On-demand single photons with high extraction efficiency and near-unity indistinguishability from a resonantly driven quantum dot in a micropillar},\ }\bibfield  {journal} {\bibinfo  {journal} {Phys.
  Rev. Lett.}\ }\textbf {\bibinfo {volume} {116}},\ \href {https://doi.org/https://doi.org/10.1103/PhysRevLett.116.020401} {https://doi.org/10.1103/PhysRevLett.116.020401} (\bibinfo {year} {2016})\BibitemShut {NoStop}%
\bibitem [{\citenamefont {Sipahigil}\ \emph {et~al.}(2012)\citenamefont {Sipahigil}, \citenamefont {Goldman}, \citenamefont {Togan}, \citenamefont {Chu}, \citenamefont {Markham}, \citenamefont {Twitchen}, \citenamefont {Zibrov}, \citenamefont {Kubanek},\ and\ \citenamefont {Lukin}}]{Sipahigil_2012}%
  \BibitemOpen
  \bibfield  {author} {\bibinfo {author} {\bibfnamefont {A.}~\bibnamefont {Sipahigil}}, \bibinfo {author} {\bibfnamefont {M.~L.}\ \bibnamefont {Goldman}}, \bibinfo {author} {\bibfnamefont {E.}~\bibnamefont {Togan}}, \bibinfo {author} {\bibfnamefont {Y.}~\bibnamefont {Chu}}, \bibinfo {author} {\bibfnamefont {M.}~\bibnamefont {Markham}}, \bibinfo {author} {\bibfnamefont {D.~J.}\ \bibnamefont {Twitchen}}, \bibinfo {author} {\bibfnamefont {A.~S.}\ \bibnamefont {Zibrov}}, \bibinfo {author} {\bibfnamefont {A.}~\bibnamefont {Kubanek}},\ and\ \bibinfo {author} {\bibfnamefont {M.~D.}\ \bibnamefont {Lukin}},\ }\bibfield  {title} {\bibinfo {title} {Quantum interference of single photons from remote nitrogen-vacancy centers in diamond},\ }\bibfield  {journal} {\bibinfo  {journal} {Phys. Rev. Lett.}\ }\textbf {\bibinfo {volume} {108}},\ \href {https://doi.org/10.1103/PhysRevLett.108.143601} {10.1103/PhysRevLett.108.143601} (\bibinfo {year} {2012})\BibitemShut {NoStop}%
\bibitem [{\citenamefont {Kumar}(1990)}]{kumar_1990}%
  \BibitemOpen
  \bibfield  {author} {\bibinfo {author} {\bibfnamefont {P.}~\bibnamefont {Kumar}},\ }\bibfield  {title} {\bibinfo {title} {Quantum frequency conversion},\ }\href {https://doi.org/10.1364/OL.15.001476} {\bibfield  {journal} {\bibinfo  {journal} {Opt. Lett.}\ }\textbf {\bibinfo {volume} {15}},\ \bibinfo {pages} {1476} (\bibinfo {year} {1990})}\BibitemShut {NoStop}%
\bibitem [{\citenamefont {Ikuta}\ \emph {et~al.}(2011)\citenamefont {Ikuta}, \citenamefont {Kusaka}, \citenamefont {Kitano}, \citenamefont {Kato}, \citenamefont {Yamamoto}, \citenamefont {Koashi},\ and\ \citenamefont {Imoto}}]{ikuta_2011}%
  \BibitemOpen
  \bibfield  {author} {\bibinfo {author} {\bibfnamefont {R.}~\bibnamefont {Ikuta}}, \bibinfo {author} {\bibfnamefont {Y.}~\bibnamefont {Kusaka}}, \bibinfo {author} {\bibfnamefont {T.}~\bibnamefont {Kitano}}, \bibinfo {author} {\bibfnamefont {H.}~\bibnamefont {Kato}}, \bibinfo {author} {\bibfnamefont {T.}~\bibnamefont {Yamamoto}}, \bibinfo {author} {\bibfnamefont {M.}~\bibnamefont {Koashi}},\ and\ \bibinfo {author} {\bibfnamefont {N.}~\bibnamefont {Imoto}},\ }\bibfield  {title} {\bibinfo {title} {Wide-band quantum interface for visible-to-telecommunication wavelength conversion},\ }\href {https://doi.org/10.1038/ncomms1544} {\bibfield  {journal} {\bibinfo  {journal} {Nat. Commun.}\ }\textbf {\bibinfo {volume} {2}},\ \bibinfo {pages} {537} (\bibinfo {year} {2011})}\BibitemShut {NoStop}%
\bibitem [{\citenamefont {Zaske}\ \emph {et~al.}(2012)\citenamefont {Zaske}, \citenamefont {Lenhard}, \citenamefont {Ke\ss{}ler}, \citenamefont {Kettler}, \citenamefont {Hepp}, \citenamefont {Arend}, \citenamefont {Albrecht}, \citenamefont {Schulz}, \citenamefont {Jetter}, \citenamefont {Michler},\ and\ \citenamefont {Becher}}]{zaske_2012}%
  \BibitemOpen
  \bibfield  {author} {\bibinfo {author} {\bibfnamefont {S.}~\bibnamefont {Zaske}}, \bibinfo {author} {\bibfnamefont {A.}~\bibnamefont {Lenhard}}, \bibinfo {author} {\bibfnamefont {C.~A.}\ \bibnamefont {Ke\ss{}ler}}, \bibinfo {author} {\bibfnamefont {J.}~\bibnamefont {Kettler}}, \bibinfo {author} {\bibfnamefont {C.}~\bibnamefont {Hepp}}, \bibinfo {author} {\bibfnamefont {C.}~\bibnamefont {Arend}}, \bibinfo {author} {\bibfnamefont {R.}~\bibnamefont {Albrecht}}, \bibinfo {author} {\bibfnamefont {W.-M.}\ \bibnamefont {Schulz}}, \bibinfo {author} {\bibfnamefont {M.}~\bibnamefont {Jetter}}, \bibinfo {author} {\bibfnamefont {P.}~\bibnamefont {Michler}},\ and\ \bibinfo {author} {\bibfnamefont {C.}~\bibnamefont {Becher}},\ }\bibfield  {title} {\bibinfo {title} {Visible-to-telecom quantum frequency conversion of light from a single quantum emitter},\ }\href {https://doi.org/10.1103/PhysRevLett.109.147404} {\bibfield  {journal} {\bibinfo  {journal} {Phys. Rev. Lett.}\ }\textbf {\bibinfo {volume} {109}},\ \bibinfo
  {pages} {147404} (\bibinfo {year} {2012})}\BibitemShut {NoStop}%
\bibitem [{\citenamefont {De~Greve}\ \emph {et~al.}(2012)\citenamefont {De~Greve}, \citenamefont {Yu}, \citenamefont {McMahon}, \citenamefont {Pelc}, \citenamefont {Natarajan}, \citenamefont {Kim}, \citenamefont {Abe}, \citenamefont {Maier}, \citenamefont {Schneider}, \citenamefont {Kamp}, \citenamefont {Höfling}, \citenamefont {Hadfield}, \citenamefont {Forchel}, \citenamefont {Fejer},\ and\ \citenamefont {Yamamoto}}]{degreve_2012}%
  \BibitemOpen
  \bibfield  {author} {\bibinfo {author} {\bibfnamefont {K.}~\bibnamefont {De~Greve}}, \bibinfo {author} {\bibfnamefont {L.}~\bibnamefont {Yu}}, \bibinfo {author} {\bibfnamefont {P.~L.}\ \bibnamefont {McMahon}}, \bibinfo {author} {\bibfnamefont {J.~S.}\ \bibnamefont {Pelc}}, \bibinfo {author} {\bibfnamefont {C.~M.}\ \bibnamefont {Natarajan}}, \bibinfo {author} {\bibfnamefont {N.~Y.}\ \bibnamefont {Kim}}, \bibinfo {author} {\bibfnamefont {E.}~\bibnamefont {Abe}}, \bibinfo {author} {\bibfnamefont {S.}~\bibnamefont {Maier}}, \bibinfo {author} {\bibfnamefont {C.}~\bibnamefont {Schneider}}, \bibinfo {author} {\bibfnamefont {M.}~\bibnamefont {Kamp}}, \bibinfo {author} {\bibfnamefont {S.}~\bibnamefont {Höfling}}, \bibinfo {author} {\bibfnamefont {R.~H.}\ \bibnamefont {Hadfield}}, \bibinfo {author} {\bibfnamefont {A.}~\bibnamefont {Forchel}}, \bibinfo {author} {\bibfnamefont {M.~M.}\ \bibnamefont {Fejer}},\ and\ \bibinfo {author} {\bibfnamefont {Y.}~\bibnamefont {Yamamoto}},\ }\bibfield  {title} {\bibinfo {title}
  {Quantum-dot spin–photon entanglement via frequency downconversion to telecom wavelength},\ }\href {https://doi.org/10.1038/nature11577} {\bibfield  {journal} {\bibinfo  {journal} {Nature}\ }\textbf {\bibinfo {volume} {491}},\ \bibinfo {pages} {421} (\bibinfo {year} {2012})}\BibitemShut {NoStop}%
\bibitem [{\citenamefont {Albrecht}\ \emph {et~al.}(2014)\citenamefont {Albrecht}, \citenamefont {Farrera}, \citenamefont {Fernandez-Gonzalvo}, \citenamefont {Cristiani},\ and\ \citenamefont {de~Riedmatten}}]{albrecht_2014}%
  \BibitemOpen
  \bibfield  {author} {\bibinfo {author} {\bibfnamefont {B.}~\bibnamefont {Albrecht}}, \bibinfo {author} {\bibfnamefont {P.}~\bibnamefont {Farrera}}, \bibinfo {author} {\bibfnamefont {X.}~\bibnamefont {Fernandez-Gonzalvo}}, \bibinfo {author} {\bibfnamefont {M.}~\bibnamefont {Cristiani}},\ and\ \bibinfo {author} {\bibfnamefont {H.}~\bibnamefont {de~Riedmatten}},\ }\bibfield  {title} {\bibinfo {title} {A waveguide frequency converter connecting rubidium-based quantum memories to the telecom {C}-band},\ }\href {https://doi.org/10.1038/ncomms4376} {\bibfield  {journal} {\bibinfo  {journal} {Nat. Commun.}\ }\textbf {\bibinfo {volume} {5}},\ \bibinfo {pages} {3376} (\bibinfo {year} {2014})}\BibitemShut {NoStop}%
\bibitem [{\citenamefont {Li}\ \emph {et~al.}(2016)\citenamefont {Li}, \citenamefont {Davanço},\ and\ \citenamefont {Srinivasan}}]{li_2016}%
  \BibitemOpen
  \bibfield  {author} {\bibinfo {author} {\bibfnamefont {Q.}~\bibnamefont {Li}}, \bibinfo {author} {\bibfnamefont {M.}~\bibnamefont {Davanço}},\ and\ \bibinfo {author} {\bibfnamefont {K.}~\bibnamefont {Srinivasan}},\ }\bibfield  {title} {\bibinfo {title} {Efficient and low-noise single-photon-level frequency conversion interfaces using silicon nanophotonics},\ }\href {https://doi.org/10.1038/nphoton.2016.64} {\bibfield  {journal} {\bibinfo  {journal} {Nat. Photonics}\ }\textbf {\bibinfo {volume} {10}},\ \bibinfo {pages} {406} (\bibinfo {year} {2016})}\BibitemShut {NoStop}%
\bibitem [{\citenamefont {Dr{\'e}au}\ \emph {et~al.}(2018)\citenamefont {Dr{\'e}au}, \citenamefont {Tchebotareva}, \citenamefont {Mahdaoui}, \citenamefont {Bonato},\ and\ \citenamefont {Hanson}}]{dreau2018quantum}%
  \BibitemOpen
  \bibfield  {author} {\bibinfo {author} {\bibfnamefont {A.}~\bibnamefont {Dr{\'e}au}}, \bibinfo {author} {\bibfnamefont {A.}~\bibnamefont {Tchebotareva}}, \bibinfo {author} {\bibfnamefont {A.~E.}\ \bibnamefont {Mahdaoui}}, \bibinfo {author} {\bibfnamefont {C.}~\bibnamefont {Bonato}},\ and\ \bibinfo {author} {\bibfnamefont {R.}~\bibnamefont {Hanson}},\ }\bibfield  {title} {\bibinfo {title} {Quantum frequency conversion of single photons from a nitrogen-vacancy center in diamond to telecommunication wavelengths},\ }\href@noop {} {\bibfield  {journal} {\bibinfo  {journal} {Physical review applied}\ }\textbf {\bibinfo {volume} {9}},\ \bibinfo {pages} {064031} (\bibinfo {year} {2018})}\BibitemShut {NoStop}%
\bibitem [{\citenamefont {Bock}\ \emph {et~al.}(2018)\citenamefont {Bock}, \citenamefont {Eich}, \citenamefont {Kucera}, \citenamefont {Kreis}, \citenamefont {Lenhard}, \citenamefont {Becher},\ and\ \citenamefont {Eschner}}]{bock2018high}%
  \BibitemOpen
  \bibfield  {author} {\bibinfo {author} {\bibfnamefont {M.}~\bibnamefont {Bock}}, \bibinfo {author} {\bibfnamefont {P.}~\bibnamefont {Eich}}, \bibinfo {author} {\bibfnamefont {S.}~\bibnamefont {Kucera}}, \bibinfo {author} {\bibfnamefont {M.}~\bibnamefont {Kreis}}, \bibinfo {author} {\bibfnamefont {A.}~\bibnamefont {Lenhard}}, \bibinfo {author} {\bibfnamefont {C.}~\bibnamefont {Becher}},\ and\ \bibinfo {author} {\bibfnamefont {J.}~\bibnamefont {Eschner}},\ }\bibfield  {title} {\bibinfo {title} {High-fidelity entanglement between a trapped ion and a telecom photon via quantum frequency conversion},\ }\href@noop {} {\bibfield  {journal} {\bibinfo  {journal} {Nature communications}\ }\textbf {\bibinfo {volume} {9}},\ \bibinfo {pages} {1998} (\bibinfo {year} {2018})}\BibitemShut {NoStop}%
\bibitem [{\citenamefont {Da~Lio}\ \emph {et~al.}(2022)\citenamefont {Da~Lio}, \citenamefont {Faurby}, \citenamefont {Zhou}, \citenamefont {Chan}, \citenamefont {Uppu}, \citenamefont {Thyrrestrup}, \citenamefont {Scholz}, \citenamefont {Wieck}, \citenamefont {Ludwig}, \citenamefont {Lodahl} \emph {et~al.}}]{da2022pure}%
  \BibitemOpen
  \bibfield  {author} {\bibinfo {author} {\bibfnamefont {B.}~\bibnamefont {Da~Lio}}, \bibinfo {author} {\bibfnamefont {C.}~\bibnamefont {Faurby}}, \bibinfo {author} {\bibfnamefont {X.}~\bibnamefont {Zhou}}, \bibinfo {author} {\bibfnamefont {M.~L.}\ \bibnamefont {Chan}}, \bibinfo {author} {\bibfnamefont {R.}~\bibnamefont {Uppu}}, \bibinfo {author} {\bibfnamefont {H.}~\bibnamefont {Thyrrestrup}}, \bibinfo {author} {\bibfnamefont {S.}~\bibnamefont {Scholz}}, \bibinfo {author} {\bibfnamefont {A.~D.}\ \bibnamefont {Wieck}}, \bibinfo {author} {\bibfnamefont {A.}~\bibnamefont {Ludwig}}, \bibinfo {author} {\bibfnamefont {P.}~\bibnamefont {Lodahl}}, \emph {et~al.},\ }\bibfield  {title} {\bibinfo {title} {A pure and indistinguishable single-photon source at telecommunication wavelength},\ }\href@noop {} {\bibfield  {journal} {\bibinfo  {journal} {Advanced Quantum Technologies}\ }\textbf {\bibinfo {volume} {5}},\ \bibinfo {pages} {2200006} (\bibinfo {year} {2022})}\BibitemShut {NoStop}%
\bibitem [{\citenamefont {Kiraz}\ \emph {et~al.}(2001)\citenamefont {Kiraz}, \citenamefont {Michler}, \citenamefont {Becher}, \citenamefont {Gayral}, \citenamefont {Imamoğlu}, \citenamefont {Zhang}, \citenamefont {Hu}, \citenamefont {Schoenfeld},\ and\ \citenamefont {Petroff}}]{kiraz_2001}%
  \BibitemOpen
  \bibfield  {author} {\bibinfo {author} {\bibfnamefont {A.}~\bibnamefont {Kiraz}}, \bibinfo {author} {\bibfnamefont {P.}~\bibnamefont {Michler}}, \bibinfo {author} {\bibfnamefont {C.}~\bibnamefont {Becher}}, \bibinfo {author} {\bibfnamefont {B.}~\bibnamefont {Gayral}}, \bibinfo {author} {\bibfnamefont {A.}~\bibnamefont {Imamoğlu}}, \bibinfo {author} {\bibfnamefont {L.}~\bibnamefont {Zhang}}, \bibinfo {author} {\bibfnamefont {E.}~\bibnamefont {Hu}}, \bibinfo {author} {\bibfnamefont {W.~V.}\ \bibnamefont {Schoenfeld}},\ and\ \bibinfo {author} {\bibfnamefont {P.~M.}\ \bibnamefont {Petroff}},\ }\bibfield  {title} {\bibinfo {title} {Cavity-quantum electrodynamics using a single inas quantum dot in a microdisk structure},\ }\href {https://doi.org/10.1063/1.1379987} {\bibfield  {journal} {\bibinfo  {journal} {Applied Physics Letters}\ }\textbf {\bibinfo {volume} {78}},\ \bibinfo {pages} {3932} (\bibinfo {year} {2001})},\ \Eprint {https://arxiv.org/abs/https://doi.org/10.1063/1.1379987}
  {https://doi.org/10.1063/1.1379987} \BibitemShut {NoStop}%
\bibitem [{\citenamefont {Faraon}\ \emph {et~al.}(2007)\citenamefont {Faraon}, \citenamefont {Englund}, \citenamefont {Fushman}, \citenamefont {Vučković}, \citenamefont {Stoltz},\ and\ \citenamefont {Petroff}}]{faraon_2007}%
  \BibitemOpen
  \bibfield  {author} {\bibinfo {author} {\bibfnamefont {A.}~\bibnamefont {Faraon}}, \bibinfo {author} {\bibfnamefont {D.}~\bibnamefont {Englund}}, \bibinfo {author} {\bibfnamefont {I.}~\bibnamefont {Fushman}}, \bibinfo {author} {\bibfnamefont {J.}~\bibnamefont {Vučković}}, \bibinfo {author} {\bibfnamefont {N.}~\bibnamefont {Stoltz}},\ and\ \bibinfo {author} {\bibfnamefont {P.}~\bibnamefont {Petroff}},\ }\bibfield  {title} {\bibinfo {title} {Local quantum dot tuning on photonic crystal chips},\ }\href {https://doi.org/10.1063/1.2742789} {\bibfield  {journal} {\bibinfo  {journal} {Applied Physics Letters}\ }\textbf {\bibinfo {volume} {90}},\ \bibinfo {pages} {213110} (\bibinfo {year} {2007})},\ \Eprint {https://arxiv.org/abs/https://doi.org/10.1063/1.2742789} {https://doi.org/10.1063/1.2742789} \BibitemShut {NoStop}%
\bibitem [{\citenamefont {Seidl}\ \emph {et~al.}(2006)\citenamefont {Seidl}, \citenamefont {Kroner}, \citenamefont {Högele}, \citenamefont {Karrai}, \citenamefont {Warburton}, \citenamefont {Badolato},\ and\ \citenamefont {Petroff}}]{seidl_2006}%
  \BibitemOpen
  \bibfield  {author} {\bibinfo {author} {\bibfnamefont {S.}~\bibnamefont {Seidl}}, \bibinfo {author} {\bibfnamefont {M.}~\bibnamefont {Kroner}}, \bibinfo {author} {\bibfnamefont {A.}~\bibnamefont {Högele}}, \bibinfo {author} {\bibfnamefont {K.}~\bibnamefont {Karrai}}, \bibinfo {author} {\bibfnamefont {R.~J.}\ \bibnamefont {Warburton}}, \bibinfo {author} {\bibfnamefont {A.}~\bibnamefont {Badolato}},\ and\ \bibinfo {author} {\bibfnamefont {P.~M.}\ \bibnamefont {Petroff}},\ }\bibfield  {title} {\bibinfo {title} {Effect of uniaxial stress on excitons in a self-assembled quantum dot},\ }\href {https://doi.org/10.1063/1.2204843} {\bibfield  {journal} {\bibinfo  {journal} {Applied Physics Letters}\ }\textbf {\bibinfo {volume} {88}},\ \bibinfo {pages} {203113} (\bibinfo {year} {2006})},\ \Eprint {https://arxiv.org/abs/https://doi.org/10.1063/1.2204843} {https://doi.org/10.1063/1.2204843} \BibitemShut {NoStop}%
\bibitem [{\citenamefont {Findeis}\ \emph {et~al.}(2001)\citenamefont {Findeis}, \citenamefont {Baier}, \citenamefont {Beham}, \citenamefont {Zrenner},\ and\ \citenamefont {Abstreiter}}]{findeis_2001}%
  \BibitemOpen
  \bibfield  {author} {\bibinfo {author} {\bibfnamefont {F.}~\bibnamefont {Findeis}}, \bibinfo {author} {\bibfnamefont {M.}~\bibnamefont {Baier}}, \bibinfo {author} {\bibfnamefont {E.}~\bibnamefont {Beham}}, \bibinfo {author} {\bibfnamefont {A.}~\bibnamefont {Zrenner}},\ and\ \bibinfo {author} {\bibfnamefont {G.}~\bibnamefont {Abstreiter}},\ }\bibfield  {title} {\bibinfo {title} {Photocurrent and photoluminescence of a single self-assembled quantum dot in electric fields},\ }\href {https://doi.org/10.1063/1.1369148} {\bibfield  {journal} {\bibinfo  {journal} {Applied Physics Letters}\ }\textbf {\bibinfo {volume} {78}},\ \bibinfo {pages} {2958} (\bibinfo {year} {2001})},\ \Eprint {https://arxiv.org/abs/https://doi.org/10.1063/1.1369148} {https://doi.org/10.1063/1.1369148} \BibitemShut {NoStop}%
\bibitem [{\citenamefont {Dubowski}\ \emph {et~al.}(2018)\citenamefont {Dubowski}, \citenamefont {Stanowski}, \citenamefont {Dalacu},\ and\ \citenamefont {Poole}}]{dubowski_2018}%
  \BibitemOpen
  \bibfield  {author} {\bibinfo {author} {\bibfnamefont {J.~J.}\ \bibnamefont {Dubowski}}, \bibinfo {author} {\bibfnamefont {R.}~\bibnamefont {Stanowski}}, \bibinfo {author} {\bibfnamefont {D.}~\bibnamefont {Dalacu}},\ and\ \bibinfo {author} {\bibfnamefont {P.~J.}\ \bibnamefont {Poole}},\ }\bibfield  {title} {\bibinfo {title} {Precision tuning of inas quantum dot emission wavelength by iterative laser annealing},\ }\href {https://doi.org/https://doi.org/10.1016/j.optlastec.2018.01.061} {\bibfield  {journal} {\bibinfo  {journal} {Optics \& Laser Technology}\ }\textbf {\bibinfo {volume} {103}},\ \bibinfo {pages} {382} (\bibinfo {year} {2018})}\BibitemShut {NoStop}%
\bibitem [{\citenamefont {Yu}\ \emph {et~al.}(2021)\citenamefont {Yu}, \citenamefont {Yuan}, \citenamefont {Zhang}, \citenamefont {Zhang}, \citenamefont {Li}, \citenamefont {Wang}, \citenamefont {Deng}, \citenamefont {You}, \citenamefont {Song}, \citenamefont {Wang}, \citenamefont {Guo},\ and\ \citenamefont {Zhou}}]{yu_2021}%
  \BibitemOpen
  \bibfield  {author} {\bibinfo {author} {\bibfnamefont {H.}~\bibnamefont {Yu}}, \bibinfo {author} {\bibfnamefont {C.}~\bibnamefont {Yuan}}, \bibinfo {author} {\bibfnamefont {R.}~\bibnamefont {Zhang}}, \bibinfo {author} {\bibfnamefont {Z.}~\bibnamefont {Zhang}}, \bibinfo {author} {\bibfnamefont {H.}~\bibnamefont {Li}}, \bibinfo {author} {\bibfnamefont {Y.}~\bibnamefont {Wang}}, \bibinfo {author} {\bibfnamefont {G.}~\bibnamefont {Deng}}, \bibinfo {author} {\bibfnamefont {L.}~\bibnamefont {You}}, \bibinfo {author} {\bibfnamefont {H.}~\bibnamefont {Song}}, \bibinfo {author} {\bibfnamefont {Z.}~\bibnamefont {Wang}}, \bibinfo {author} {\bibfnamefont {G.-C.}\ \bibnamefont {Guo}},\ and\ \bibinfo {author} {\bibfnamefont {Q.}~\bibnamefont {Zhou}},\ }\href@noop {} {\bibinfo {title} {Spectrally multiplexed heralded single photon source at telecom-band}} (\bibinfo {year} {2021}),\ \Eprint {https://arxiv.org/abs/2104.02593} {arXiv:2104.02593 [quant-ph]} \BibitemShut {NoStop}%
\bibitem [{\citenamefont {Grimau~Puigibert}\ \emph {et~al.}(2017)\citenamefont {Grimau~Puigibert}, \citenamefont {Aguilar}, \citenamefont {Zhou}, \citenamefont {Marsili}, \citenamefont {Shaw}, \citenamefont {Verma}, \citenamefont {Nam}, \citenamefont {Oblak},\ and\ \citenamefont {Tittel}}]{puigibert_2017}%
  \BibitemOpen
  \bibfield  {author} {\bibinfo {author} {\bibfnamefont {M.}~\bibnamefont {Grimau~Puigibert}}, \bibinfo {author} {\bibfnamefont {G.~H.}\ \bibnamefont {Aguilar}}, \bibinfo {author} {\bibfnamefont {Q.}~\bibnamefont {Zhou}}, \bibinfo {author} {\bibfnamefont {F.}~\bibnamefont {Marsili}}, \bibinfo {author} {\bibfnamefont {M.~D.}\ \bibnamefont {Shaw}}, \bibinfo {author} {\bibfnamefont {V.~B.}\ \bibnamefont {Verma}}, \bibinfo {author} {\bibfnamefont {S.~W.}\ \bibnamefont {Nam}}, \bibinfo {author} {\bibfnamefont {D.}~\bibnamefont {Oblak}},\ and\ \bibinfo {author} {\bibfnamefont {W.}~\bibnamefont {Tittel}},\ }\bibfield  {title} {\bibinfo {title} {Heralded single photons based on spectral multiplexing and feed-forward control},\ }\href {https://doi.org/10.1103/PhysRevLett.119.083601} {\bibfield  {journal} {\bibinfo  {journal} {Phys. Rev. Lett.}\ }\textbf {\bibinfo {volume} {119}},\ \bibinfo {pages} {083601} (\bibinfo {year} {2017})}\BibitemShut {NoStop}%
\bibitem [{\citenamefont {Nobuyuki}(2016)}]{matsuda_2016}%
  \BibitemOpen
  \bibfield  {author} {\bibinfo {author} {\bibfnamefont {M.}~\bibnamefont {Nobuyuki}},\ }\bibfield  {title} {\bibinfo {title} {Deterministic reshaping of single-photon spectra using cross-phase modulation},\ }\href {https://doi.org/10.1126/sciadv.1501223} {\bibfield  {journal} {\bibinfo  {journal} {Science Advances}\ }\textbf {\bibinfo {volume} {2}},\ \bibinfo {pages} {e1501223} (\bibinfo {year} {2016})},\ \Eprint {https://arxiv.org/abs/https://www.science.org/doi/pdf/10.1126/sciadv.1501223} {https://www.science.org/doi/pdf/10.1126/sciadv.1501223} \BibitemShut {NoStop}%
\bibitem [{\citenamefont {Takesue}(2008)}]{takesue_2008}%
  \BibitemOpen
  \bibfield  {author} {\bibinfo {author} {\bibfnamefont {H.}~\bibnamefont {Takesue}},\ }\bibfield  {title} {\bibinfo {title} {Erasing distinguishability using quantum frequency up-conversion},\ }\href {https://doi.org/10.1103/PhysRevLett.101.173901} {\bibfield  {journal} {\bibinfo  {journal} {Phys. Rev. Lett.}\ }\textbf {\bibinfo {volume} {101}},\ \bibinfo {pages} {173901} (\bibinfo {year} {2008})}\BibitemShut {NoStop}%
\bibitem [{\citenamefont {Ates}\ \emph {et~al.}(2012)\citenamefont {Ates}, \citenamefont {Agha}, \citenamefont {Gulinatti}, \citenamefont {Rech}, \citenamefont {Rakher}, \citenamefont {Badolato},\ and\ \citenamefont {Srinivasan}}]{ates_2012}%
  \BibitemOpen
  \bibfield  {author} {\bibinfo {author} {\bibfnamefont {S.}~\bibnamefont {Ates}}, \bibinfo {author} {\bibfnamefont {I.}~\bibnamefont {Agha}}, \bibinfo {author} {\bibfnamefont {A.}~\bibnamefont {Gulinatti}}, \bibinfo {author} {\bibfnamefont {I.}~\bibnamefont {Rech}}, \bibinfo {author} {\bibfnamefont {M.~T.}\ \bibnamefont {Rakher}}, \bibinfo {author} {\bibfnamefont {A.}~\bibnamefont {Badolato}},\ and\ \bibinfo {author} {\bibfnamefont {K.}~\bibnamefont {Srinivasan}},\ }\bibfield  {title} {\bibinfo {title} {Two-photon interference using background-free quantum frequency conversion of single photons emitted by an {InAs} quantum dot},\ }\href {https://doi.org/10.1103/PhysRevLett.109.147405} {\bibfield  {journal} {\bibinfo  {journal} {Phys. Rev. Lett.}\ }\textbf {\bibinfo {volume} {109}},\ \bibinfo {pages} {147405} (\bibinfo {year} {2012})}\BibitemShut {NoStop}%
\bibitem [{\citenamefont {Weber}\ \emph {et~al.}(2019)\citenamefont {Weber}, \citenamefont {Kambs}, \citenamefont {Kettler}, \citenamefont {Kern}, \citenamefont {Maisch}, \citenamefont {Vural}, \citenamefont {Jetter}, \citenamefont {Portalupi}, \citenamefont {Becher},\ and\ \citenamefont {Michler}}]{weber_2019}%
  \BibitemOpen
  \bibfield  {author} {\bibinfo {author} {\bibfnamefont {J.~H.}\ \bibnamefont {Weber}}, \bibinfo {author} {\bibfnamefont {B.}~\bibnamefont {Kambs}}, \bibinfo {author} {\bibfnamefont {J.}~\bibnamefont {Kettler}}, \bibinfo {author} {\bibfnamefont {S.}~\bibnamefont {Kern}}, \bibinfo {author} {\bibfnamefont {J.}~\bibnamefont {Maisch}}, \bibinfo {author} {\bibfnamefont {H.}~\bibnamefont {Vural}}, \bibinfo {author} {\bibfnamefont {M.}~\bibnamefont {Jetter}}, \bibinfo {author} {\bibfnamefont {S.~L.}\ \bibnamefont {Portalupi}}, \bibinfo {author} {\bibfnamefont {C.}~\bibnamefont {Becher}},\ and\ \bibinfo {author} {\bibfnamefont {P.}~\bibnamefont {Michler}},\ }\bibfield  {title} {\bibinfo {title} {Two-photon interference in the telecom {C}-band after frequency conversion of photons from remote quantum emitters},\ }\href {https://doi.org/10.1038/s41565-018-0279-8} {\bibfield  {journal} {\bibinfo  {journal} {Nat. Nanotechnol.}\ }\textbf {\bibinfo {volume} {14}},\ \bibinfo {pages} {23} (\bibinfo {year}
  {2019})}\BibitemShut {NoStop}%
\bibitem [{\citenamefont {Singh}\ \emph {et~al.}(2019)\citenamefont {Singh}, \citenamefont {Li}, \citenamefont {Liu}, \citenamefont {Yu}, \citenamefont {Lu}, \citenamefont {Schneider}, \citenamefont {H\"{o}fling}, \citenamefont {Lawall}, \citenamefont {Verma}, \citenamefont {Mirin}, \citenamefont {Nam}, \citenamefont {Liu},\ and\ \citenamefont {Srinivasan}}]{singh_2019}%
  \BibitemOpen
  \bibfield  {author} {\bibinfo {author} {\bibfnamefont {A.}~\bibnamefont {Singh}}, \bibinfo {author} {\bibfnamefont {Q.}~\bibnamefont {Li}}, \bibinfo {author} {\bibfnamefont {S.}~\bibnamefont {Liu}}, \bibinfo {author} {\bibfnamefont {Y.}~\bibnamefont {Yu}}, \bibinfo {author} {\bibfnamefont {X.}~\bibnamefont {Lu}}, \bibinfo {author} {\bibfnamefont {C.}~\bibnamefont {Schneider}}, \bibinfo {author} {\bibfnamefont {S.}~\bibnamefont {H\"{o}fling}}, \bibinfo {author} {\bibfnamefont {J.}~\bibnamefont {Lawall}}, \bibinfo {author} {\bibfnamefont {V.}~\bibnamefont {Verma}}, \bibinfo {author} {\bibfnamefont {R.}~\bibnamefont {Mirin}}, \bibinfo {author} {\bibfnamefont {S.~W.}\ \bibnamefont {Nam}}, \bibinfo {author} {\bibfnamefont {J.}~\bibnamefont {Liu}},\ and\ \bibinfo {author} {\bibfnamefont {K.}~\bibnamefont {Srinivasan}},\ }\bibfield  {title} {\bibinfo {title} {Quantum frequency conversion of a quantum dot single-photon source on a nanophotonic chip},\ }\href {https://doi.org/10.1364/OPTICA.6.000563} {\bibfield
  {journal} {\bibinfo  {journal} {Optica}\ }\textbf {\bibinfo {volume} {6}},\ \bibinfo {pages} {563} (\bibinfo {year} {2019})}\BibitemShut {NoStop}%
\bibitem [{\citenamefont {Eckstein}\ \emph {et~al.}(2011)\citenamefont {Eckstein}, \citenamefont {Brecht},\ and\ \citenamefont {Silberhorn}}]{eckstein_2011}%
  \BibitemOpen
  \bibfield  {author} {\bibinfo {author} {\bibfnamefont {A.}~\bibnamefont {Eckstein}}, \bibinfo {author} {\bibfnamefont {B.}~\bibnamefont {Brecht}},\ and\ \bibinfo {author} {\bibfnamefont {C.}~\bibnamefont {Silberhorn}},\ }\bibfield  {title} {\bibinfo {title} {A quantum pulse gate based on spectrally engineered sum frequency generation},\ }\href {https://doi.org/10.1364/OE.19.013770} {\bibfield  {journal} {\bibinfo  {journal} {Opt. Express}\ }\textbf {\bibinfo {volume} {19}},\ \bibinfo {pages} {13770} (\bibinfo {year} {2011})}\BibitemShut {NoStop}%
\bibitem [{\citenamefont {Brecht}\ \emph {et~al.}(2011)\citenamefont {Brecht}, \citenamefont {Eckstein}, \citenamefont {Christ}, \citenamefont {Suche},\ and\ \citenamefont {Silberhorn}}]{brecht_2011}%
  \BibitemOpen
  \bibfield  {author} {\bibinfo {author} {\bibfnamefont {B.}~\bibnamefont {Brecht}}, \bibinfo {author} {\bibfnamefont {A.}~\bibnamefont {Eckstein}}, \bibinfo {author} {\bibfnamefont {A.}~\bibnamefont {Christ}}, \bibinfo {author} {\bibfnamefont {H.}~\bibnamefont {Suche}},\ and\ \bibinfo {author} {\bibfnamefont {C.}~\bibnamefont {Silberhorn}},\ }\bibfield  {title} {\bibinfo {title} {From quantum pulse gate to quantum pulse shaper{\textemdash}engineered frequency conversion in nonlinear optical waveguides},\ }\href {https://doi.org/10.1088/1367-2630/13/6/065029} {\bibfield  {journal} {\bibinfo  {journal} {New Journal of Physics}\ }\textbf {\bibinfo {volume} {13}},\ \bibinfo {pages} {065029} (\bibinfo {year} {2011})}\BibitemShut {NoStop}%
\bibitem [{\citenamefont {Brecht}\ \emph {et~al.}(2014)\citenamefont {Brecht}, \citenamefont {Eckstein}, \citenamefont {Ricken}, \citenamefont {Quiring}, \citenamefont {Suche}, \citenamefont {Sansoni},\ and\ \citenamefont {Silberhorn}}]{brecht_2014}%
  \BibitemOpen
  \bibfield  {author} {\bibinfo {author} {\bibfnamefont {B.}~\bibnamefont {Brecht}}, \bibinfo {author} {\bibfnamefont {A.}~\bibnamefont {Eckstein}}, \bibinfo {author} {\bibfnamefont {R.}~\bibnamefont {Ricken}}, \bibinfo {author} {\bibfnamefont {V.}~\bibnamefont {Quiring}}, \bibinfo {author} {\bibfnamefont {H.}~\bibnamefont {Suche}}, \bibinfo {author} {\bibfnamefont {L.}~\bibnamefont {Sansoni}},\ and\ \bibinfo {author} {\bibfnamefont {C.}~\bibnamefont {Silberhorn}},\ }\bibfield  {title} {\bibinfo {title} {Demonstration of coherent time-frequency schmidt mode selection using dispersion-engineered frequency conversion},\ }\href {https://doi.org/10.1103/PhysRevA.90.030302} {\bibfield  {journal} {\bibinfo  {journal} {Phys. Rev. A}\ }\textbf {\bibinfo {volume} {90}},\ \bibinfo {pages} {030302} (\bibinfo {year} {2014})}\BibitemShut {NoStop}%
\bibitem [{\citenamefont {Brecht}\ \emph {et~al.}(2015)\citenamefont {Brecht}, \citenamefont {Reddy}, \citenamefont {Silberhorn},\ and\ \citenamefont {Raymer}}]{brecht_2015}%
  \BibitemOpen
  \bibfield  {author} {\bibinfo {author} {\bibfnamefont {B.}~\bibnamefont {Brecht}}, \bibinfo {author} {\bibfnamefont {D.~V.}\ \bibnamefont {Reddy}}, \bibinfo {author} {\bibfnamefont {C.}~\bibnamefont {Silberhorn}},\ and\ \bibinfo {author} {\bibfnamefont {M.~G.}\ \bibnamefont {Raymer}},\ }\bibfield  {title} {\bibinfo {title} {Photon temporal modes: A complete framework for quantum information science},\ }\href {https://doi.org/10.1103/PhysRevX.5.041017} {\bibfield  {journal} {\bibinfo  {journal} {Phys. Rev. X}\ }\textbf {\bibinfo {volume} {5}},\ \bibinfo {pages} {041017} (\bibinfo {year} {2015})}\BibitemShut {NoStop}%
\bibitem [{\citenamefont {Grice}\ and\ \citenamefont {Walmsley}(1997)}]{grice_1997}%
  \BibitemOpen
  \bibfield  {author} {\bibinfo {author} {\bibfnamefont {W.~P.}\ \bibnamefont {Grice}}\ and\ \bibinfo {author} {\bibfnamefont {I.~A.}\ \bibnamefont {Walmsley}},\ }\bibfield  {title} {\bibinfo {title} {Spectral information and distinguishability in type-ii down-conversion with a broadband pump},\ }\href {https://doi.org/10.1103/PhysRevA.56.1627} {\bibfield  {journal} {\bibinfo  {journal} {Phys. Rev. A}\ }\textbf {\bibinfo {volume} {56}},\ \bibinfo {pages} {1627} (\bibinfo {year} {1997})}\BibitemShut {NoStop}%
\bibitem [{\citenamefont {Quesada}\ and\ \citenamefont {Sipe}(2014)}]{quesada2014effects}%
  \BibitemOpen
  \bibfield  {author} {\bibinfo {author} {\bibfnamefont {N.}~\bibnamefont {Quesada}}\ and\ \bibinfo {author} {\bibfnamefont {J.}~\bibnamefont {Sipe}},\ }\bibfield  {title} {\bibinfo {title} {Effects of time ordering in quantum nonlinear optics},\ }\href@noop {} {\bibfield  {journal} {\bibinfo  {journal} {Physical Review A}\ }\textbf {\bibinfo {volume} {90}},\ \bibinfo {pages} {063840} (\bibinfo {year} {2014})}\BibitemShut {NoStop}%
\bibitem [{\citenamefont {Manzoni}\ and\ \citenamefont {Cerullo}(2016)}]{Manzoni2016}%
  \BibitemOpen
  \bibfield  {author} {\bibinfo {author} {\bibfnamefont {C.}~\bibnamefont {Manzoni}}\ and\ \bibinfo {author} {\bibfnamefont {G.}~\bibnamefont {Cerullo}},\ }\bibfield  {title} {\bibinfo {title} {Design criteria for ultrafast optical parametric amplifiers},\ }\href {https://doi.org/10.1088/2040-8978/18/10/103501} {\bibfield  {journal} {\bibinfo  {journal} {Journal of Optics}\ }\textbf {\bibinfo {volume} {18}},\ \bibinfo {pages} {103501} (\bibinfo {year} {2016})}\BibitemShut {NoStop}%
\bibitem [{\citenamefont {Graffitti}\ \emph {et~al.}(2018)\citenamefont {Graffitti}, \citenamefont {Kelly-Massicotte}, \citenamefont {Fedrizzi},\ and\ \citenamefont {Bra\ifmmode~\acute{n}\else \'{n}\fi{}czyk}}]{Graffitti2018}%
  \BibitemOpen
  \bibfield  {author} {\bibinfo {author} {\bibfnamefont {F.}~\bibnamefont {Graffitti}}, \bibinfo {author} {\bibfnamefont {J.}~\bibnamefont {Kelly-Massicotte}}, \bibinfo {author} {\bibfnamefont {A.}~\bibnamefont {Fedrizzi}},\ and\ \bibinfo {author} {\bibfnamefont {A.~M.}\ \bibnamefont {Bra\ifmmode~\acute{n}\else \'{n}\fi{}czyk}},\ }\bibfield  {title} {\bibinfo {title} {Design considerations for high-purity heralded single-photon sources},\ }\href {https://doi.org/10.1103/PhysRevA.98.053811} {\bibfield  {journal} {\bibinfo  {journal} {Phys. Rev. A}\ }\textbf {\bibinfo {volume} {98}},\ \bibinfo {pages} {053811} (\bibinfo {year} {2018})}\BibitemShut {NoStop}%
\bibitem [{\citenamefont {Law}\ \emph {et~al.}(2000)\citenamefont {Law}, \citenamefont {Walmsley},\ and\ \citenamefont {Eberly}}]{law_2000}%
  \BibitemOpen
  \bibfield  {author} {\bibinfo {author} {\bibfnamefont {C.~K.}\ \bibnamefont {Law}}, \bibinfo {author} {\bibfnamefont {I.~A.}\ \bibnamefont {Walmsley}},\ and\ \bibinfo {author} {\bibfnamefont {J.~H.}\ \bibnamefont {Eberly}},\ }\bibfield  {title} {\bibinfo {title} {Continuous frequency entanglement: Effective finite {H}ilbert space and entropy control},\ }\href {https://doi.org/10.1103/PhysRevLett.84.5304} {\bibfield  {journal} {\bibinfo  {journal} {Phys. Rev. Lett.}\ }\textbf {\bibinfo {volume} {84}},\ \bibinfo {pages} {5304} (\bibinfo {year} {2000})}\BibitemShut {NoStop}%
\bibitem [{\citenamefont {Parker}\ \emph {et~al.}(2000)\citenamefont {Parker}, \citenamefont {Bose},\ and\ \citenamefont {Plenio}}]{parker_2000}%
  \BibitemOpen
  \bibfield  {author} {\bibinfo {author} {\bibfnamefont {S.}~\bibnamefont {Parker}}, \bibinfo {author} {\bibfnamefont {S.}~\bibnamefont {Bose}},\ and\ \bibinfo {author} {\bibfnamefont {M.~B.}\ \bibnamefont {Plenio}},\ }\bibfield  {title} {\bibinfo {title} {Entanglement quantification and purification in continuous-variable systems},\ }\href {https://doi.org/10.1103/PhysRevA.61.032305} {\bibfield  {journal} {\bibinfo  {journal} {Phys. Rev. A}\ }\textbf {\bibinfo {volume} {61}},\ \bibinfo {pages} {032305} (\bibinfo {year} {2000})}\BibitemShut {NoStop}%
\bibitem [{\citenamefont {Law}\ and\ \citenamefont {Eberly}(2004)}]{law_2004}%
  \BibitemOpen
  \bibfield  {author} {\bibinfo {author} {\bibfnamefont {C.~K.}\ \bibnamefont {Law}}\ and\ \bibinfo {author} {\bibfnamefont {J.~H.}\ \bibnamefont {Eberly}},\ }\bibfield  {title} {\bibinfo {title} {Analysis and interpretation of high transverse entanglement in optical parametric down conversion},\ }\href {https://doi.org/10.1103/PhysRevLett.92.127903} {\bibfield  {journal} {\bibinfo  {journal} {Phys. Rev. Lett.}\ }\textbf {\bibinfo {volume} {92}},\ \bibinfo {pages} {127903} (\bibinfo {year} {2004})}\BibitemShut {NoStop}%
\bibitem [{\citenamefont {Pelc}\ \emph {et~al.}(2011)\citenamefont {Pelc}, \citenamefont {Ma}, \citenamefont {Phillips}, \citenamefont {Zhang}, \citenamefont {Langrock}, \citenamefont {Slattery}, \citenamefont {Tang},\ and\ \citenamefont {Fejer}}]{pelc2011long}%
  \BibitemOpen
  \bibfield  {author} {\bibinfo {author} {\bibfnamefont {J.~S.}\ \bibnamefont {Pelc}}, \bibinfo {author} {\bibfnamefont {L.}~\bibnamefont {Ma}}, \bibinfo {author} {\bibfnamefont {C.}~\bibnamefont {Phillips}}, \bibinfo {author} {\bibfnamefont {Q.}~\bibnamefont {Zhang}}, \bibinfo {author} {\bibfnamefont {C.}~\bibnamefont {Langrock}}, \bibinfo {author} {\bibfnamefont {O.}~\bibnamefont {Slattery}}, \bibinfo {author} {\bibfnamefont {X.}~\bibnamefont {Tang}},\ and\ \bibinfo {author} {\bibfnamefont {M.~M.}\ \bibnamefont {Fejer}},\ }\bibfield  {title} {\bibinfo {title} {Long-wavelength-pumped upconversion single-photon detector at 1550 nm: performance and noise analysis},\ }\href@noop {} {\bibfield  {journal} {\bibinfo  {journal} {Optics express}\ }\textbf {\bibinfo {volume} {19}},\ \bibinfo {pages} {21445} (\bibinfo {year} {2011})}\BibitemShut {NoStop}%
\bibitem [{\citenamefont {Esfandyarpour}\ \emph {et~al.}(2018)\citenamefont {Esfandyarpour}, \citenamefont {Langrock},\ and\ \citenamefont {Fejer}}]{esfandyarpour2018cascaded}%
  \BibitemOpen
  \bibfield  {author} {\bibinfo {author} {\bibfnamefont {V.}~\bibnamefont {Esfandyarpour}}, \bibinfo {author} {\bibfnamefont {C.}~\bibnamefont {Langrock}},\ and\ \bibinfo {author} {\bibfnamefont {M.}~\bibnamefont {Fejer}},\ }\bibfield  {title} {\bibinfo {title} {Cascaded downconversion interface to convert single-photon-level signals at 650 nm to the telecom band},\ }\href@noop {} {\bibfield  {journal} {\bibinfo  {journal} {Optics Letters}\ }\textbf {\bibinfo {volume} {43}},\ \bibinfo {pages} {5655} (\bibinfo {year} {2018})}\BibitemShut {NoStop}%
\bibitem [{\citenamefont {Geus}\ \emph {et~al.}(2024)\citenamefont {Geus}, \citenamefont {Elsen}, \citenamefont {Nyga}, \citenamefont {Stolk}, \citenamefont {van~der Enden}, \citenamefont {van Zwet}, \citenamefont {Haefner}, \citenamefont {Hanson},\ and\ \citenamefont {Jungbluth}}]{geus2024low}%
  \BibitemOpen
  \bibfield  {author} {\bibinfo {author} {\bibfnamefont {J.~F.}\ \bibnamefont {Geus}}, \bibinfo {author} {\bibfnamefont {F.}~\bibnamefont {Elsen}}, \bibinfo {author} {\bibfnamefont {S.}~\bibnamefont {Nyga}}, \bibinfo {author} {\bibfnamefont {A.~J.}\ \bibnamefont {Stolk}}, \bibinfo {author} {\bibfnamefont {K.~L.}\ \bibnamefont {van~der Enden}}, \bibinfo {author} {\bibfnamefont {E.~J.}\ \bibnamefont {van Zwet}}, \bibinfo {author} {\bibfnamefont {C.}~\bibnamefont {Haefner}}, \bibinfo {author} {\bibfnamefont {R.}~\bibnamefont {Hanson}},\ and\ \bibinfo {author} {\bibfnamefont {B.}~\bibnamefont {Jungbluth}},\ }\bibfield  {title} {\bibinfo {title} {Low-noise short-wavelength pumped frequency downconversion for quantum frequency converters},\ }\href@noop {} {\bibfield  {journal} {\bibinfo  {journal} {Optica Quantum}\ }\textbf {\bibinfo {volume} {2}},\ \bibinfo {pages} {189} (\bibinfo {year} {2024})}\BibitemShut {NoStop}%
\bibitem [{\citenamefont {Schneeloch}\ \emph {et~al.}(2019)\citenamefont {Schneeloch}, \citenamefont {Knarr}, \citenamefont {Bogorin}, \citenamefont {Levangie}, \citenamefont {Tison}, \citenamefont {Frank}, \citenamefont {Howland}, \citenamefont {Fanto},\ and\ \citenamefont {Alsing}}]{Schneeloch_2019}%
  \BibitemOpen
  \bibfield  {author} {\bibinfo {author} {\bibfnamefont {J.}~\bibnamefont {Schneeloch}}, \bibinfo {author} {\bibfnamefont {S.~H.}\ \bibnamefont {Knarr}}, \bibinfo {author} {\bibfnamefont {D.~F.}\ \bibnamefont {Bogorin}}, \bibinfo {author} {\bibfnamefont {M.~L.}\ \bibnamefont {Levangie}}, \bibinfo {author} {\bibfnamefont {C.~C.}\ \bibnamefont {Tison}}, \bibinfo {author} {\bibfnamefont {R.}~\bibnamefont {Frank}}, \bibinfo {author} {\bibfnamefont {G.~A.}\ \bibnamefont {Howland}}, \bibinfo {author} {\bibfnamefont {M.~L.}\ \bibnamefont {Fanto}},\ and\ \bibinfo {author} {\bibfnamefont {P.~M.}\ \bibnamefont {Alsing}},\ }\bibfield  {title} {\bibinfo {title} {Introduction to the absolute brightness and number statistics in spontaneous parametric down-conversion},\ }\href {https://doi.org/10.1088/2040-8986/ab05a8} {\bibfield  {journal} {\bibinfo  {journal} {Journal of Optics}\ }\textbf {\bibinfo {volume} {21}},\ \bibinfo {pages} {043501} (\bibinfo {year} {2019})}\BibitemShut {NoStop}%
\bibitem [{\citenamefont {Tamo\v{s}auskas}\ \emph {et~al.}(2018)\citenamefont {Tamo\v{s}auskas}, \citenamefont {Beresnevi\v{c}ius}, \citenamefont {Gadonas},\ and\ \citenamefont {Dubietis}}]{tamosauskas_2018}%
  \BibitemOpen
  \bibfield  {author} {\bibinfo {author} {\bibfnamefont {G.}~\bibnamefont {Tamo\v{s}auskas}}, \bibinfo {author} {\bibfnamefont {G.}~\bibnamefont {Beresnevi\v{c}ius}}, \bibinfo {author} {\bibfnamefont {D.}~\bibnamefont {Gadonas}},\ and\ \bibinfo {author} {\bibfnamefont {A.}~\bibnamefont {Dubietis}},\ }\bibfield  {title} {\bibinfo {title} {Transmittance and phase matching of {BBO} crystal in the 3--5 \textmu m range and its application for the characterization of mid-infrared laser pulses},\ }\href {https://doi.org/10.1364/OME.8.001410} {\bibfield  {journal} {\bibinfo  {journal} {Opt. Mater. Express}\ }\textbf {\bibinfo {volume} {8}},\ \bibinfo {pages} {1410} (\bibinfo {year} {2018})}\BibitemShut {NoStop}%
\bibitem [{\citenamefont {Kato}(1986)}]{kato_1986}%
  \BibitemOpen
  \bibfield  {author} {\bibinfo {author} {\bibfnamefont {K.}~\bibnamefont {Kato}},\ }\bibfield  {title} {\bibinfo {title} {Second-harmonic generation to 2048 {$\mathrm{\AA}$} in {$\beta$-Ba$_2$O$_4$}},\ }\href {https://doi.org/10.1109/JQE.1986.1073097} {\bibfield  {journal} {\bibinfo  {journal} {IEEE J. Quantum Electron.}\ }\textbf {\bibinfo {volume} {22}},\ \bibinfo {pages} {1013} (\bibinfo {year} {1986})}\BibitemShut {NoStop}%
\bibitem [{\citenamefont {Umemura}\ \emph {et~al.}(2007)\citenamefont {Umemura}, \citenamefont {Miyata},\ and\ \citenamefont {Kato}}]{umemura_2007}%
  \BibitemOpen
  \bibfield  {author} {\bibinfo {author} {\bibfnamefont {N.}~\bibnamefont {Umemura}}, \bibinfo {author} {\bibfnamefont {K.}~\bibnamefont {Miyata}},\ and\ \bibinfo {author} {\bibfnamefont {K.}~\bibnamefont {Kato}},\ }\bibfield  {title} {\bibinfo {title} {New data on the optical properties of {BiB$_3$O$_6$}},\ }\href {https://doi.org/https://doi.org/10.1016/j.optmat.2006.12.014} {\bibfield  {journal} {\bibinfo  {journal} {Opt. Mater.}\ }\textbf {\bibinfo {volume} {30}},\ \bibinfo {pages} {532} (\bibinfo {year} {2007})}\BibitemShut {NoStop}%
\bibitem [{\citenamefont {Kato}(1994)}]{kato_1994}%
  \BibitemOpen
  \bibfield  {author} {\bibinfo {author} {\bibfnamefont {K.}~\bibnamefont {Kato}},\ }\bibfield  {title} {\bibinfo {title} {Temperature-tuned 90$^\circ$ phase-matching properties of {LiB$_3$O$_5$}},\ }\href {https://doi.org/10.1109/3.362711} {\bibfield  {journal} {\bibinfo  {journal} {IEEE J. Quantum Electron.}\ }\textbf {\bibinfo {volume} {30}},\ \bibinfo {pages} {2950} (\bibinfo {year} {1994})}\BibitemShut {NoStop}%
\bibitem [{\citenamefont {Kato}(1985)}]{kato_1985}%
  \BibitemOpen
  \bibfield  {author} {\bibinfo {author} {\bibfnamefont {K.}~\bibnamefont {Kato}},\ }\bibfield  {title} {\bibinfo {title} {High-power difference-frequency generation at 4.4--5.7 µm in {LiIO$_3$}},\ }\href {https://doi.org/10.1109/JQE.1985.1072617} {\bibfield  {journal} {\bibinfo  {journal} {IEEE J. Quantum Electron.}\ }\textbf {\bibinfo {volume} {21}},\ \bibinfo {pages} {119} (\bibinfo {year} {1985})}\BibitemShut {NoStop}%
\bibitem [{\citenamefont {Kato}\ and\ \citenamefont {Takaoka}(2002)}]{kato_2002}%
  \BibitemOpen
  \bibfield  {author} {\bibinfo {author} {\bibfnamefont {K.}~\bibnamefont {Kato}}\ and\ \bibinfo {author} {\bibfnamefont {E.}~\bibnamefont {Takaoka}},\ }\bibfield  {title} {\bibinfo {title} {Sellmeier and thermo-optic dispersion formulas for {KTP}},\ }\href {https://doi.org/10.1364/AO.41.005040} {\bibfield  {journal} {\bibinfo  {journal} {Appl. Opt.}\ }\textbf {\bibinfo {volume} {41}},\ \bibinfo {pages} {5040} (\bibinfo {year} {2002})}\BibitemShut {NoStop}%
\bibitem [{\citenamefont {Zelmon}\ \emph {et~al.}(1997)\citenamefont {Zelmon}, \citenamefont {Small},\ and\ \citenamefont {Jundt}}]{zelmon_1997}%
  \BibitemOpen
  \bibfield  {author} {\bibinfo {author} {\bibfnamefont {D.~E.}\ \bibnamefont {Zelmon}}, \bibinfo {author} {\bibfnamefont {D.~L.}\ \bibnamefont {Small}},\ and\ \bibinfo {author} {\bibfnamefont {D.}~\bibnamefont {Jundt}},\ }\bibfield  {title} {\bibinfo {title} {Infrared corrected {S}ellmeier coefficients for congruently grown lithium niobate and 5 mol. \% magnesium oxide-doped niobate},\ }\href {https://doi.org/10.1364/JOSAB.14.003319} {\bibfield  {journal} {\bibinfo  {journal} {J. Opt. Soc. Am. B}\ }\textbf {\bibinfo {volume} {14}},\ \bibinfo {pages} {3319} (\bibinfo {year} {1997})}\BibitemShut {NoStop}%
\end{thebibliography}%

\end{document}